\documentstyle[aps,prb,floats,psfig]{revtex}
\begin{document}
%\draft
%\tightenlines
%\twocolumn[\hsize\textwidth\columnwidth\hsize\csname@twocolumnfalse\endcsname
\title{Quantum Decoherence and Weak Localization at Low Temperatures}
\author{Dmitrii S. Golubev$^{1,3}$ and Andrei D. Zaikin$^{2,3}$}
\address{$^{1}$ Physics Department, Chalmers University of Technology,
S-41296 G\"oteborg, Sweden\\
$^2$ Institut f\"{u}r Theoretische Festk\"orperphysik,
Universit\"at Karlsruhe, 76128 Karlsruhe, Germany\\
$^3$ I.E.Tamm Department of Theoretical Physics, P.N.Lebedev
Physics Institute, Leninskii pr. 53, 117924 Moscow, Russia}

\maketitle

\begin{abstract}
We develope a theory of a fundamental effect of the interaction-induced
decoherence of the electron wave function in a disordered metal.
With the aid of the Keldysh technique and the path integral formalism
we derive a formally exact equation of motion for the electron
density matrix in the presence of interaction. We demonstrate that
the effect of interaction of the electron with other electrons and
lattice ions in a disordered metal is equivalent to that of an effective
dissipative environment. Quantum noise of this environment causes
quantum decoherence even at $T=0$. Our analysis explicitely accounts
for the Pauli principle which plays an important role for inelastic
scattering processes but turns out not to affect quantum decoherence.
Our results seriously challenge the existence of strong localization
in low dimensional disordered metals.
\end{abstract}


%\begin{multicols}{2}
%]

\section{Introduction}

Recent experiments \cite{Webb} attracted a great deal of attention
to an old but fundamental question: how fast can a quantum particle
loose information about its initial state in the presence of interaction?
In other words, how fast can interaction destroy the quantum phase
coherence? The answer on this question essentially depends on the type
of interaction.

It follows from general principles of quantum
mechanics that quantum coherence of the wave function cannot
be destroyed due to {\it elastic} interaction with an external
potential. Another physical situation may take place if the quantum particle
interacts with other (quantum) degrees of freedom which play the
role of an effective environment. In this case quantum dynamics of
the particle cannot be described by the wave function but
only by the density matrix. Various examples (to be discussed below) show
that such interactions {\it may lead} to a complete destruction of
quantum coherence.

A general approach to the problem
was formulated by Feynman and Vernon \cite{FV,FH} who demonstrated
that the effect of environment can be taken into account by
means of averaging over its all possible quantum states. As a
result the environment variables are integrated out
and quantum dynamics of the particle can be described only
in terms of its own degrees of freedom. Within this approach
interaction
with the external environment is taken into account by
means of the so-called {\it influence functional} which
appears in the (effective) action for the particle as a
result of averaging
over the bath variables. It is quite clear that specific
properties of the environment are not important unless
they explicitely enter the expression for the influence functional.
In other words, the particle does not ``feel'' the difference
between physically different bathes provided they are
described by the same influence functional.

These ideas were developed further by Caldeira and Leggett
\cite{CL} who showed that the above arguments can be used
to describe quantum dynamics of dissipative systems and
derived the effective action for the
case of linear Ohmic dissipation. The same type of analysis
was also developed by Schmid \cite{Schmid1} who formulated
a quasiclassical Langevin equation approach describing
real time dynamics of a quantum particle in the presence
of dissipation and quantum noise.

Although the above papers are dealing with the model of
a bosonic environment it is obvious that the ideas \cite{FV,FH}
can be applied to a fermionic bath as well. This was done
e.g. by Ambegaokar, Eckern and Sch\"on \cite{AES} in the case of
superconducting tunnel junctions and later by Sch\"on and one of
the present authors \cite{SZ89,SZ} in a somewhat broader context
of a metallic system with dissipation. Although the microscopic
Hamiltonian describing electrons in a metal \cite{AES,SZ89,SZ}
is quite different from one used in the model \cite{CL} the final
expressions for the influence functionals for various
metallic systems and tunnel junctions obtained in \cite{AES,SZ89,SZ}
turn out to be similar or even completely equivalent to those
considered in Refs. \onlinecite{CL,Schmid1}. This equivalence
is just an illustration of the property discussed above:
the effect of physically different environments is indistinguishable
provided they are described by the same influence functional.

In Refs. \onlinecite{AES,SZ89,SZ} quantum dynamics of a certain collective
variable of interest (the phase) was considered. This variable
was extracted ``from interaction'',
after that electronic degrees of freedom were integrated out and the
effective action for the phase was derived. One can also
generalize this procedure and describe quantum dynamics of superconductors
considering the phase as a quantum field \cite{OGZB}. In all
these cases the collective variable is intimately linked to the
electronic bath, quantum dynamics of the former does not exist without
the latter at all.

In this paper we will analyze a somewhat different situation. Namely,
we will study quantum dynamics of an electron propagating
in a disordered metal and interacting with other electrons
which play the role of effective environment. It is well known
that quantum interference of electrons scattered on impurities
lead to quantum corrections to the classical Drude conductivity
\cite{AAR,GLK}. These so-called weak localization corrections
have been extensively discussed in the literature (see e.g. \cite{AAK1,CS}
for review). The magnitude of these corrections is known to
be determined by the time within which electrons in a metal
can be described by a phase coherent wave function. At times
exceeding this so-called decoherence time $\tau_{\varphi}$
quantum coherence is destroyed, quantum interference is not anymore
possible and therefore the classical diffusion picture is restored.

The decoherence time $\tau_{\varphi}$ in a disordered metal can
be determined by various physical processes, such as
electron-electron and  electron-phonon interactions, electron
scattering on magnetic impurities etc. (see e.g. \cite{AAK1,CS}).
It was shown by Altshuler, Aronov and Khmelnitskii \cite{AAK}
that at not very low temperatures the effect of electron-electron
interaction on the decoherence time is equivalent to that of
classical Nyquist noise in a disordered conductor. In this case
one finds \cite{AAK,CS,SAI,Imry}
\begin{equation}
\tau_{\varphi} \propto T^{2/(d-4)},
\label{AAK}
\end{equation}
where $d$ is the effective system dimension. This result
demonstrates that decoherence effect of electron-electron interaction
becomes weaker as the temperature is lowered.

Down to which temperature does the above result remain correct?
Or, more generally, does $\tau_{\varphi}$ increase with decreasing
$T$ at all temperatures thus going to infinity at $T \to 0$?
A positive answer on the latter question would mean that at $T=0$
the electron in a disordered metal can be described
by the phase coherent wave function even in the presence of
Coulomb interaction with other electrons. This scenario faces
several important problems:

(i) Even if one assumes that electrons in a metal are isolated from
the external environment it is hard to believe that the $N$-electron
wave function $\Psi_N$ ($N=10^a$ is the total number of electrons in a
conductor, $a \gg 1$) can be factorized into the product of
single electron wave functions. This would imply that Coulomb
interaction is effectively renormalized to zero at $T=0$. We see
no reason to expect that.

(ii) In any real physical situation electrons in a metal {\it are}
coupled to the external environment, e.g. via external leads. Electrons
can enter and leave a piece of metal, interact with electrons in the
leads etc. Due to this reason the whole system of $N$ electrons in a
metal (and not only one electron!) should be treated as open and thus
cannot be described by the wave function $\Psi_N$ even at $T=0$.

(iii) In the case of Caldeira-Leggett type of models it is
well known that after a sufficienly long time even at $T=0$
quantum coherence is destroyed due to interaction with the
bath no matter how weak this interaction is (see e.g. \cite{ChLeg}).
For strong enough interaction the particle exhibits practically
no quantum features and can even get localized in a potential well
\cite{CBM,Schmid2}. Since the Caldeira-Leggett models are
qualitatively similar to the problem in question (in both cases
one quantum degree of freedom interacts with many others playing
the role of a bath) it is quite natural to expect also the
decoherence time for interacting electrons in a disordered
metal to remain finite even at $T=0$.

(iv) Saturation of the temperature dependence for $\tau_{\varphi}$
at low $T$ was clearly observed in experiments \cite{Webb}
as well as in numerous earlier experimental studies (see e.g.
\cite{Mikko,Gio,Pooke,Hir,Mueller}). Even a qualitative interpretation
of these results in terms of eq. (\ref{AAK}) is not possible.
There exist sufficiently strong experimental reasons to believe that
this saturation is {\it not} due to heating of the sample or
the effect of magnetic impurities.

All the above arguments strongly suggest that the decoherence
time $\tau_{\varphi}$ should remain finite down to $T=0$. Although
the experimental evidence for the low temperature saturation
of $\tau_{\varphi}$ is available for quite a long time the
physical reason for this effect was not convincingly explained
in the literature. An earlier attempt \cite{Kumar} to provide
this explanation in terms of zero-point fluctuations of
impurities has failed \cite{Berg}.
Only very recently it was suggested \cite{Webb,GZ97} that
quantum decoherence in a disordered conductor is caused by
the mechanism qualitatively similar to one existing in the
Caldeira-Leggett model, namely the interaction of a particle with
an external quantum bath. In the present case an electron
propagating in a disordered metal interacts with a fluctuating
electric field produced by other electrons which play the role
of an effective {\it dissipative} environment. Due to
randomness of such fluctuations the quantum coherence is lost
within a finite time $\tau_{\varphi}$.

In fact, this mechanism is not new. Exactly the same effect was
considered in Ref. \onlinecite{AAK} where only the low frequency (classical)
fluctuations of the environment $\omega <T$ were taken into
account. This approximation is sufficient at high temperatures
and yields the result (\ref{AAK}). However, as the temperature
is lowered there remain less and less classical modes in
the fluctuating environment. Therefore it is reasonable to
expect that below a certain temperature $T_q$ quantum fluctuations
of the environment with $\omega >T$ will take over and determine
the decoherence time $\tau_{\varphi}$. The generalization of
the corresponding analysis \cite{AAK,CS,SAI,Imry} is straightforward
\cite{GZ97} and allows to evaluate both the classical-to-quantum
crossover temperature $T_q$ and the decoherence time
$\tau_{\varphi}$ at low $T$. The results \cite{GZ97} agree quite
well with the experimental findings \cite{Webb}.

Several fundamentally important questions should still be
answered, however.
Perhaps the main one is about the role of the Pauli principle
in our description of the electron-electron interaction.
At sufficiently high temperatures there are always enough
empty states into which electrons can get scattered
provided the energy transfer due to interaction does not
exceed $T$. In this case the Pauli principle does not play
a very important role. On the other hand, at $T \to 0$ in
equilibrium all quantum states below the Fermi energy are
occupied. Thus scattering into any of these states is
forbidden due to the Pauli principle and the electron
energy cannot decrease. At $T=0$ it also cannot increase
due to interaction with other electrons because they are also
in the ground state. Thus at $T=0$ inelastic electron-electron
scattering is forbidden due to the Pauli principle. If so,
how can one expect quantum decoherence to occur as a result
of the electron-electron interaction at $T=0$? One might think
that due to the Pauli principle the electron is simply
insensitive to zero-point fluctuations of other electrons,
and quantum decoherence does not take place.

Another though somewhat related question is about the effect of
phonons. It is quite obvious that at a finite $T$ decoherence can
also be caused by
electron-phonon interaction \cite{AAK1,CS}. However at $T\to 0$ there are no
real phonons in the system and therefore no quantum decoherence
due to phonons should be expected. On the other hand, if we
admit that zero-point fluctuations of electrons lead to decoherence,
zero point fluctuations of the lattice ions should have qualitatively
the same effect. Is it possible to match this conclusion e.g. with the
well known fact that electrons do not scatter on zero-point lattice
fluctuations, and therefore the phonon contribution to the conductance
of a metal vanishes at $T=0$?

On a more formal level it is sometimes conjectured that the above arguments
correspond to the exact cancellation of Keldysh diagrams for $\tau_{\varphi}$
in the limit $T\to 0$. If
this is indeed the case the inverse decoherence time $1/\tau_{\varphi}$
would be exactly equal to zero at $T=0$ and (at least qualitatively)
would match with the result (\ref{AAK}).

The main goal of this paper is to demonstrate that this is {\it not} the case.

The key point is that all the above arguments apply as far as
the electron {\it inelastic relaxation time} $\tau_i$ is concerned.
The latter can be found by means of a quantum kinetic approach
which will be fully reproduced within the framework of our analysis. However,
as it has been pointed out in Ref. \onlinecite{AAK} and discussed
later in other works \cite{Blanter,Imry,GZ97}, the decoherence time
$\tau_{\varphi}$ is entirely different from $\tau_i$ and, in contrast
to the latter, cannot be determined with the aid of any kind of kinetic
analysis. Roughly speaking, the decoherence time controls the decay of
off-diagonal elements of the electron density matrix and (even at
high temperatures) has little to do with inelastic scattering
of electrons. Therefore the quasiclassical kinetic treatment is
insufficient and a fully quantum mechanical analysis should be
elaborated to determine the time $\tau_{\varphi}$.

The paper is organized as follows. In Section 2 we make use
of the general formalism of the Green-Keldysh functions and derive
a formally exact equation of motion for the electron density matrix in
the presence of Coulomb interaction. This equation explicitely accounts
for the Pauli principle and allows for a clear understanding of
its role in the process of electro-electron interaction in a metal.
In Section 3 the effective action (or the influence functional)
for the fluctuating scalar potential in a metallic conductor is
derived. We also demonstrate that in equilibrium this influence
functional satisfies the fluctuation-dissipation theorem \cite{LL}
and establish the relation with the real time effective action
derived in the Caldeira-Leggett models \cite{CL,Schmid1,SZ,GZ92}.
In Section 4 with the aid of these general results we will
derive the real time effective action for the electron propagating
in a metal and determine the decoherence time $\tau_{\varphi}$
and the weak localization correction to conductivity at low temperatures.
Our formalism naturally includes both electron-electron and
electron-phonon interactions and allows to establish the corresponding
contributions to $\tau_{\varphi}$ from each of these processes.
In Section 5 we derive the quasiclassical kinetic equation and demonstrate
the relation of our analysis to the standard kinetic approach which
allows to evaluate the inelastic scattering time $\tau_i$. We also
derive the quasiclassical Langevin equation which under certain conditions
can be used to describe propagation of electrons in a disordered metal.
Discussion of the results is given in Section 6. We discuss
the relation of our results with previous works as well as possible
consequences of the low temperature saturation of $\tau_{\varphi}$
for the existing picture of strong localization in low dimensional
conductors. We also briefly compare our theoretical predictions with
available experimental data. Some details of our calculation are presented in
Appendices.

\section{Density Matrix}
We will consider a standard Hamiltonian describing electrons in a disordered
metal
\begin{equation}
\bbox{H}_{el}=\bbox{H}_0+\bbox{H}_{int},
\end{equation}
where
\begin{equation}
\bbox{H}_0= \int d\bbox{r}\psi^+_{\sigma}(\bbox{r})
\left[-\frac{\nabla^2}{2m}-\mu +U(\bbox{r})\right]\psi_\sigma (\bbox{r}),
\label{H0}
\end{equation}
\begin{equation}
\bbox{H}_{int}=\frac{1}{2}
\int d\bbox{r}\int d\bbox{r'}\psi^+_{\sigma}(\bbox{r})\psi^+_{\sigma'}(\bbox{r'})
e^2v(\bbox{r}-\bbox{r'})\psi_{\sigma'}(\bbox{r'})\psi_{\sigma}(\bbox{r}).
\label{Hint}
\end{equation}
Here $\mu$ is the chemical potential, $U(\bbox{r})$ accounts for a random
potential due to nonmagnetic impurities, and
$v(\bbox{r})=1/|\bbox{r}|$ represents
the Coulomb interaction between electrons.

Let us define the generating functional for the electron Green-Keldysh
functions in terms of the path integral over the Grassman fields $\bar \psi$
and $\psi$
\begin{equation}
J[\eta ,\eta^* ]=\frac{\int{\cal D}V\int{\cal D}\bar\psi\int{\cal D}\psi
\exp \big(iS_{eff}[\bar\psi ,\psi , V] +
i\int_{K} dt\int d\bbox{r}[\bar\psi (t,\bbox{r})\eta (t,\bbox{r})+
\psi (t,\bbox{r})\eta^* (t,\bbox{r}]\big)}{\int{\cal D}V\int{\cal D}\bar\psi
\int{\cal D}\psi
\exp \big(iS_{eff}[\bar\psi ,\psi , V]\big)},
\label{genfunc}
\end{equation}
where $S_{eff}$ is the effective action
$$
S_{eff}[\bar\psi ,\psi , V]=\int_{K} dt\left(\int d\bbox{r}
[i\bar\psi (t,\bbox{r})\partial_t\psi (t,\bbox{r})-
e\bar\psi (t,\bbox{r})\psi (t,\bbox{r})V(t,\bbox{r})]-\bbox{H}_0[\bar\psi ,\psi]
\right)
$$
\begin{equation}
+\frac12\int_{K} dt\int d\bbox{r}\int d\bbox{r'}
V(t,\bbox{r})v^{-1}(\bbox{r}-\bbox{r'})V(t,\bbox{r'}),
\label{Seff}
\end{equation}
where $v^{-1}(\bbox{r}-\bbox{r'})=-\nabla^2/4\pi$.
Integration over time $t$ in (\ref{Seff}) goes along the Keldysh
contour $K$ which runs in the forward and then in the backward time
directions \cite{Keldysh}. In (\ref{genfunc},\ref{Seff}) we performed a standard
Hubbard-Stratonovich transformation introducing the path integral
over a scalar potential field $V$ in order to decouple the $\psi^4$-interaction
in (\ref{Hint}). The electron Green-Keldysh function $\bbox{\hat G}$ can be
determined from (\ref{genfunc}) by taking the derivatives
with respect to the source fields $\eta$ and $\eta^*$:
\begin{equation}
\bbox{\hat G}(t,\bbox{r};t',\bbox{r'})=i\frac{\delta}{\delta \eta^* (t,\bbox{r})}
\frac{\delta}{\delta \eta (t',\bbox{r'})}J[\eta ,\eta^* ]|_{\eta =\eta^*=0}.
\label{GK}
\end{equation}
Making use of (\ref{genfunc}, \ref{GK}) and the definition of the Green-Keldysh
function for an electron interacting with the field $V$
\begin{equation}
\hat G_V(t,\bbox{r};t',\bbox{r'})=-i \frac{\int{\cal D}\bar\psi\int{\cal D}\psi
\:
\psi (t, \bbox{r})\bar \psi (t',\bbox{r'})\exp (iS_{eff}[\bar\psi ,\psi , V])}
{\int{\cal D}\bar\psi\int{\cal D}\psi\exp (iS_{eff}[\bar\psi ,\psi , V])}
\label{GKV}
\end{equation}
it is easy to prove the identity
\begin{equation}
\bbox{\hat G}=\frac{\int{\cal D}V_1{\cal D}V_2
\; \hat G_V\; e^{i\bbox{S}[V_1,V_2]}}
{\int{\cal D}V_1{\cal D}V_2\; e^{iS[V_1,V_2]}},
\label{G1}
\end{equation}
where
\begin{equation}
i\bbox{S}[V_1,V_2]=2{\rm Tr}\ln\hat G_V^{-1}+ i\int\limits_0^t dt'\int
d\bbox{r} \frac{(\nabla V_1)^2-(\nabla V_2)^2}{8\pi}.
\label{lndet}
\end{equation}
The factor 2 in front of the trace comes from the summation
over a spin index. In (\ref{G1},\ref{lndet}) we explicitely defined the
fields $V_1(t)$ and $V_2(t)$ equal to $V(t)$ respectively on the forward
and backward parts on the Keldysh contour $K$.
Analogously $\bbox{\hat G}$ and $\hat G_V\equiv\hat G[V_1,V_2]$ are the
2$\times$2 matrices in the Keldysh space:
\begin{equation}
\bbox{\hat G}=\left( \begin{array}{cc} \bbox{G}_{11} & -\bbox{G}_{12} \\
\bbox{G}_{21} & -\bbox{G}_{22} \end{array} \right).
\label{bfG}
\end{equation}
The matrix function $\hat G_V$ obeys the equation
\begin{equation}
\left(i\frac{\partial}{\partial t_1}-\hat
H_0(\bbox{r}_1)+e\hat V(t_1,\bbox{r}_1)\right) \hat G_V=\delta(t_1-t_2)
\delta(\bbox{r}_1-\bbox{r}_2);
\label{Schrodinger}
\end{equation}
where
\begin{equation}
\hat H_0=H_0\hat 1=\left( \begin{array}{c}
                \frac{-\nabla^2}{2m}-\mu+U(\bbox{r}) \qquad 0 \\
                0 \qquad \frac{-\nabla^2}{2m}-\mu+U(\bbox{r})
                \end{array} \right);\;\;\;\;\;\;\;
\hat V=\left( \begin{array}{cc}
                V_1(t,\bbox{r}) & 0 \\
                0    & V_2(t,\bbox{r})
                \end{array} \right).
\label{V}
\end{equation}
Note that the function $\hat G_V$ is to some extent similar to the Green-Keldysh
function of an electron in an external field. However there exists
an important difference: in our case the electron interacts with a fluctuating
(quantum) field $V$. Formally this implies that the fields $V(t,\bbox{r})$ on
two parts of the Keldysh contour differ
$V_1(t,\bbox{r})\neq V_2(t,\bbox{r})$ while for the external field one always
has $V_1(t,\bbox{r})\equiv V_2(t,\bbox{r})$.

The general solution of the equation (\ref{Schrodinger})
can be expressed in the form
\begin{equation}
\hat G_V(t_1,t_2) = -i\hat U_V(t_1,t_2)[\theta(t_1-t_2)\hat a
-\theta(t_2-t_1) \hat b +\hat f_V(t_2)],
\label{Uf}
\end{equation}
Here we defined
\begin{equation}
\hat a=\left( \begin{array}{cc}
          1 & 0 \\
          0 & 0
       \end{array} \right), \qquad
\hat b=\left( \begin{array}{cc}
          0 & 0 \\
          0 & 1
       \end{array} \right),
\end{equation}
$\hat U_V(t_1,t_2)$ is the matrix evolution operator
\begin{equation}
\hat U_V(t_1,t_2) = \left( \begin{array}{cc}
                          u_1(t_1,t_2) & 0 \\
                          0 & u_2(t_1,t_2)
                         \end{array} \right),
\label{hatU}
\end{equation}
which consists of the scalar evolution operators
\begin{eqnarray}
u_{1,2}(t_1,t_2)&=&{\bf T}\exp\left[-i\int\limits_{t_1}^{t_2} dt'
(H_0-eV_{1,2}(t'))\right]
\nonumber \\
&=&
\int\limits_{\bbox{r}(t_1)=\bbox{r}_i}^{\bbox{r}(t_2)=\bbox{r}_f} {\cal D}
\bbox{r}(t')
\exp\left[i\int\limits_{t_1}^{t_2} dt'
\left(\frac{m\bbox{\dot r}^2}{2}-U(\bbox{r})+eV_{1,2}(t',\bbox{r})\right)\right],
\label{u}
\end{eqnarray}
{\bf T} is the time ordering operator. In eq. (\ref{Uf}) and below we always
imply integration over the internal coordinate variables in the product of
operators, whereas integration over time is written explicitely. For the sake
of brevity we also do not indicate the coordinate dependence in (\ref{Uf})
and many subsequent expressions. This dependence can be trivially restored
if needed.

Note that eq. (\ref{Uf}) is completely equivalent to
the standard representation of the Green-Keldysh matrix which elements
can be expressed in terms of retarded, advanced and Keldysh Green
functions. The representation (\ref{Uf}) defines a general solution of the
linear differential equation (\ref{Schrodinger}):
the term $\hat U_V(t_1,t_2)\hat f_V(t_2)$ with an arbitrary matrix operator
$\hat f_V(t_2)$ represents a general solution of the homogeneous equation,
while the terms with $\theta$-functions give a particular solution of the
inhomogeneous equation.
The operator function $\hat f_V(t_2)$ in (\ref{Uf}) is fixed by the Dyson equation
\begin{equation}
\hat G_V(t_1,t_2)=\hat G_0(t_1,t_2) -\int\limits_0^t dt'
\hat G_0(t_1,t') e\hat V(t')\hat G_V(t',t_2).
\label{Dyson}
\end{equation}
The matrix $\hat G_0$ is the electron Green-Keldysh function
without the field. This function is defined by eqs.
(\ref{Uf}-\ref{u}) with $V_{1,2}(t,\bbox{r})\equiv 0$ and
$\hat f_0(t_2)$ has the form
\begin{equation}
\hat f_0(t_2)=\left( \begin{array}{cc}
                     -\rho_0(t_2)  & \rho_0(t_2) \\
                    1-\rho_0(t_2)  & \rho_0(t_2)
                    \end{array}\right),
\label{f0}
\end{equation}
where $\rho_0(t)=e^{-iH_0t}\rho(0)e^{iH_0t}$ is the electron
density matrix for $V_{1,2}=0$ at a time $t$.

The equation (\ref{Dyson})
can be solved perturbatively in $e\hat V$. Combining this solution
with eq. (\ref{G1}) one reproduces the standard Keldysh diagrams. This way of
treating the problem is quite complicated in general and becomes
particularly nontransparent in the interesting limit of low temperatures.

We will proceed differently.

It is well known that the 1,2-component of the Green-Keldysh matrix $\bbox{\hat G}$
is directly related to the exact electron density matrix
\begin{equation}
\bbox{\rho} (t; \bbox{r},\bbox{r'})=-i\bbox{G}_{12}(t,t;\bbox{r},\bbox{r'}),
\label{rhoG}
\end{equation}
which contains all necessary information about the system dynamics
in the presence of interaction. Analogously one can define the ``density
matrix'' $\rho_V(t)\equiv f_{1,2}(t)$ related to the 1,2-component
of the matrix $\hat G_V$ by the equation equivalent to (\ref{rhoG}).
Our strategy is as follows. First we will derive the exact equation
of motion for the density matrix $\rho_V(t)$ which turns out to have a
very simple and transparent form. Already at this stage we will
clarify the role of the fluctuating fields $V_{1,2}$ and the
Pauli principle in our problem. Then we will evaluate the influence
functional $S[V_1,V_2]$ (\ref{lndet}) and find the density matrix $\bbox{\rho}$
from the equation
\begin{equation}
\bbox{\rho} (t; \bbox{r},\bbox{r'})=
\langle \rho_V (t; \bbox{r},\bbox{r'}\rangle_{V_1,V_2},
\label{rhorho}
\end{equation}
where the average over fields $V_1$ and $V_2$ is defined in (\ref{G1}).

The derivation of the equation for the density matrix $\rho_V(t)$ is
straightforward. Let us perform the time integration in the last term
of eq. (\ref{Dyson}). Integrating by parts and making use of eq.
(\ref{Schrodinger}) after a simple algebra (see Appendix A) we obtain
\begin {equation}
\hat G_0(t_1,t)\hat G(t,t_2) - \hat G_0(t_1,0)\hat G(0,t_2) = 0.
\label{diG}
\end{equation}
Substituting the representation (\ref{Uf}) into (\ref{diG}) we arrive
at the matrix equation which relates the matrix $\hat f_V(t)$, the
evolution operator $\hat U_V(t)$ and the initial density matrix $\rho_0(0)$
defined for $V_{1,2}=0$. With the aid of this equation one determines
the 1,2-component of the matrix $\hat f_V(t)$ and thus the density matrix
$\rho_V(t)$. The details of this calculation are presented in Appendix A.
As a result we find
\begin{equation}
[1-\rho_0(t)(u_2(t,0)u_1(0,t)-1)]\rho_V(t)=\rho_0(t).
\label{rhot}
\end{equation}
One can also rewrite this result in the form of the
differential equation describing the time evolution of the density matrix:
\begin{equation}
i\frac{\partial \rho_V}{\partial t} =
[H_0,\rho_V] - (1-\rho_V)eV_1\rho_V + \rho_V eV_2 (1-\rho_V),
\qquad \rho_V(0)=\rho_0(0).
\label{rho10}
\end{equation}
The equation (\ref{rho10}) is the main result of this Section. We would
like to emphasize that our derivation was performed {\it without any approximation}, i.e. the result (\ref{rho10}) {\it is exact}. It contains all
information about the system dynamics hidden in the four components of the
Green-Keldysh matrix. In the absence of the fluctuating field $V_{1,2}=0$ eq.
(\ref{rho10}) reduces to the standard equation for the electron density
matrix with the Hamiltonian $H_0$. In the presence of the field $V_{1,2}$
the equation (\ref{rho10}) exactly accounts for the Pauli principle. This is
obvious from our derivation which automatically takes care about the
Fermi statistics through the integration over the Grassman fields
$\psi$ and $\bar \psi$. This is also quite clear from the form
of the last two terms in the right hand side of eq. (\ref{rho10}). In Section 5
we will demonstrate that within the quantum kinetic analysis these terms
are responsible for the standard in- and out-scattering terms in the collision
integral.

In order to understand the role of the Pauli principle let us
rewrite the equation (\ref{rho10}) in the form
\begin{equation}
i\frac{\partial\rho_V}{\partial t}=
[H_0-eV^+,\rho_V] - (1-\rho_V)\frac{eV^-}{2}\rho_V -
\rho_V\frac{eV^-}{2}(1-\rho_V),
\label{rho5}
\end{equation}
where we defined $V^+=(V_1+V_2)/2$ and $V^-=V_1-V_2$.
It is quite obvious from (\ref{rho5}) that the field
$V^+(t,\bbox{r})$ plays the same role as an external field. All electrons
move collectively in this field, its presence is equivalent to local
fluctuations of the Fermi energy $\mu \to \mu +eV^+(t,\bbox{r})$.
The Pauli principle does not play any role here.
Below we will demonstrate that quantum fluctuations of the field $V^+$
are responsible for the low temperature saturation of the decoherence
time $\tau_{\varphi}$ and the weak localization correction to conductivity
in disordered metals.

The field $V^-$ is, on the contrary, very sensitive
to the Pauli principle. It will be shown below that this field
is responsible for damping due to radiation of an electron which
moves in a metal. Corresponding energy losses can be only due to
electron transitions into lower energy states. At $T=0$ in equilibrium
all such states are already occupied by other electrons, therefore
such processes are forbidden and the electron energy remains unchanged
due to the Pauli principle. We will demonstrate, however, that these
processes are irrelevant for the decay of the off-diagonal elements
of the electron density matrix and therefore the Pauli principle can
hardly affect quantum decoherence in a disordered metal even at $T=0$.

\section{Influence Functional for the Field}

Let us now derive the expression for the influence functional (effective action)
$S[V_1,V_2]$ for the field $V$. A formally exact action $\bbox{S}[V_1,V_2]$
obtained by integration over all electron degrees of freedom
is given by eq. (\ref{lndet}). Let us expand this expression up to the second
order in $V_{1,2}$. The first order order terms of this expansion vanish
because the Green function $\hat G_0$ corresponds to a
zero current and zero charge density state of the system.
In the second order we obtain
\begin{eqnarray}
i\delta S^{(2)}&=&-{\rm Tr}(\hat G_0e\hat V \hat G_0e\hat V)
\nonumber \\
&=& -e^2{\rm tr}\bigg[
G_{11}V^+G_{11}V^+ - G_{12}V^+G_{21}V^+
- G_{21}V^+G_{12}V^+ + G_{22}V^+G_{22}V^+
\nonumber \\
&&
G_{11}V^+G_{11}V^- - G_{12}V^+G_{21}V^-
+ G_{21}V^+G_{12}V^- - G_{22}V^+G_{22}V^-
\nonumber \\
&&
\frac{1}{4}G_{11}V^-G_{11}V^- +\frac{1}{4} G_{12}V^-G_{21}V^-
+\frac{1}{4} G_{21}V^-G_{12}V^- +\frac{1}{4} G_{22}V^-G_{22}V^-
\bigg]
\label{2order}
\end{eqnarray}
Here the subscript $_0$ for the Green functions is omitted for the sake of
simplicity, all the Green functions here and below in this Section
are defined for $V_{1,2}=0$. The expression (\ref{2order}) can be
simplified with the aid of the identities
\begin{equation}
G_{11}=G_{12} + G^R =G_{21}+G^A, \qquad
G_{22}=G_{12} - G^A= G_{21}-G^R,
\label{identity}
\end{equation}
which
allow to exclude the functions $G_{11}$ and $G_{22}$ from the action
(\ref{2order}). The terms containing $V^+V^+$ are reduced to
\begin{equation}
{\rm tr}(G^RV^+G^RV^+ + G^AV^+G^AV^+).
\label{V++}
\end{equation}
Making use of the expressions
\begin{equation}
G^R=-i\theta(t_1-t_2)u_0(t_1,t_2), \qquad G^A=i\theta(t_2-t_1)u_0(t_1,t_2),
\label{GRA}
\end{equation}
and writing the traces (\ref{V++}) in the time-space representation,
we immediately observe that the product $\theta(t_1-t_2)\theta(t_2-t_1)=0$
appears under the integral, and the whole combination (\ref{V++}) vanish.
In other words, the terms $V^+V^+$ give no contribution to the
action (\ref{2order}). The remaining terms in combination with the
last two terms in (\ref{lndet}) describing the free field action yield
%\begin{eqnarray}
%iS&=&-e^2{\rm tr}\bigg[
%G^RV^+(G_{21}+G_{12})V^- + (G_{12}+G_{21})V^+G^AV^-
%\nonumber \\
%&&
%+\frac{1}{2}G_{12}V^-G_{21}V^- + \frac{1}{2}G_{21}V^-G_{12}V^- \bigg]
%\nonumber \\
%&&
%+i\int\limits_0^t dt'\int d^3x \frac{(\nabla V^+\nabla V^-)}{4\pi}
%\label{action}
%\end{eqnarray}

\begin{eqnarray}
iS[V_1,V_2]&=&
i\int\limits_0^t dt'\int d \bbox{r} \frac{(\nabla V^+\nabla V^-)}{4\pi}
-\int\limits_0^t dt_1\int\limits_0^t dt_2
\int d\bbox{r}_1 d\bbox{r}_2\times
\nonumber \\
&&
\bigg[
iV^-(t_1,\bbox{r}_1)\chi(t_1,t_2,\bbox{r}_1,\bbox{r}_2)V^+(t_2,\bbox{r}_2)+
V^-(t_1,\bbox{r}_1)\eta(t_1,t_2,\bbox{r}_1,\bbox{r}_2)V^-(t_2,\bbox{r}_2)
\bigg],
\label{act}
\end{eqnarray}
where
\begin{eqnarray}
\chi(t_1,t_2,\bbox{r}_1,\bbox{r}_2)=
-ie^2\big[G^R(t_1,t_2,\bbox{r}_1,\bbox{r}_2)\big(G_{21}(t_2,t_1,\bbox{r}_2,
\bbox{r}_1)+G_{12}(t_2,t_1,\bbox{r}_2,\bbox{r}_1)\big)+
\nonumber \\
\big(G_{12}(t_1,t_2,\bbox{r}_1,\bbox{r}_2)+G_{21}(t_1,t_2,\bbox{r}_1,\bbox{r}_2)
\big)G^A(t_2,t_1,\bbox{r}_2,\bbox{r}_1)
\big]
\label{chi1}
\end{eqnarray}
\begin{eqnarray}
\eta(t_1,t_2,\bbox{r}_1,\bbox{r}_2)=
\frac{e^2}{2}\big[G_{12}(t_1,t_2,\bbox{r}_1,\bbox{r}_2)
G_{21}(t_2,t_1,\bbox{r}_2,\bbox{r}_1)+
G_{21}(t_1,t_2,\bbox{r}_1,\bbox{r}_2)G_{12}(t_2,t_1,\bbox{r}_2,\bbox{r}_1)\big]
\label{eta1}
\end{eqnarray}

The expressions (\ref{act}-\ref{eta1}) define the influence functional for
the field $V$ in terms of the Green-Keldysh functions for the (in general
nonequilibrium) electron subsystem. It is easy to check that the above
expressions satisfy the requirement of causality: in the
$V^-(t_1)V^+(t_2)$ terms the time $t_1$ is always larger than $t_2$ due
to analytic properties of retarded and advanced Green functions (\ref{GRA}).
It is also straightforward to demonstrate (see Appendix B) that in
thermodynamic equilibrium the kernels $\eta(\omega)$ and
${\rm Im}\chi(\omega)$ (Fourier transformed with respect
to the time difference $t_1-t_2$) satisfy the equation
\begin{equation}
\eta(\omega,\bbox{r}_1,\bbox{r}_2)=
-\frac{1}{2}\coth\left(\frac{\omega}{2T}\right)
{\rm Im}\chi(\omega,\bbox{r}_1,\bbox{r}_2).
\label{FDT}
\end{equation}
The latter equation is just the fluctuation-dissipation theorem \cite{LL}.

Finally, for a homogeneous system one can also perform the Fourier transformation
with respect to $\bbox{r}_1-\bbox{r}_2$ and find
\begin{eqnarray}
iS[V_1,V_2]=
i\int\frac{d\omega d^3k}{(2\pi)^4}
V^-(-\omega,-k)\frac{k^2\epsilon(\omega,k)}{4\pi}V^+(\omega,k)-
\nonumber \\
-\frac{1}{2}
\int\frac{d\omega d^3k}{(2\pi)^4}
V^-(-\omega,-k)\frac{k^2{\rm Im}\epsilon(\omega,k)}{4\pi}
\coth\left(\frac{\omega}{2T}\right)
V^-(\omega,k);
\label{actionf}
\end{eqnarray}
where $\epsilon (\omega , k)$ is the dielectric susceptibility of the system. For a
homogeneous electron gas it is given by the standard RPA formula:
\begin{equation}
\epsilon(\omega,k)=1+\frac{e^2}{\pi^2k^2}
\int d^3p \frac{n_{p+k}-n_p}{\omega-\xi_{p+k}+\xi_p+i0}.
\label{epsilon}
\end{equation}

Eq. (\ref{epsilon}) includes only the electron contribution to the
susceptibility. In general the effect of ions should also be accounted for.
Here we will describe this effect within a very simple approximation which is,
however, sufficient for our analysis. Namely, the ion contribution to the
susceptibility will be taken in the form
$\delta\epsilon_i=-\omega_{pi}^2/\omega^2$, where $\omega_{pi}$ is the ion
plasma frequency. Then the phonon spectrum is determined
by the equation $\epsilon(\omega,k)=0$.
In the long wave limit this approximation works sufficiently well for
longitudinal phonons which mainly interact with electrons.
The effect of transverse phonons cannot be described within this simple model.
But such phonons are weakly coupled to the elecrons anyway, and therefore
their effect can be safely ignored. If needed, further generalizations
of this simple model can be also incorporated into our analysis.

In the relevant case of a disordered metal a direct
calculation of polarization
bubbles (see e.g. \cite{OGZB}) yields
\begin{equation}
\epsilon(\omega,k)=1+\frac{4\pi\sigma}{-i\omega +Dk^2}
-\frac{\omega_{pi}^2}{\omega^2} .
\label{eps}
\end{equation}
Here $\sigma =2e^2N_0D$ is the classical Drude conductivity, $N_0$ is the
metallic density of states and $D=v_Fl/3$ is the diffusion coefficient.

The expression (\ref{eps}) is valid for wave vectors smaller
than the inverse elastic mean free path $k\lesssim 1/l$ and for small
frequencies $\omega\lesssim 1/\tau_e$, where $\tau_e=l/v_F$ is the
elastic mean free time. Note, that
if one neglects the effect of phonons and considers only nearly uniform in
space ($k \approx 0$) fluctuations of the field $V$ one immediately
observes that eqs. (\ref{actionf},\ref{eps}) {\it exactly} coincide with
the real time version of the Caldeira-Leggett action
\cite{CL,Schmid1,SZ89,SZ} in this limit.
For a disordered metal this action was derived by means of the
quasiclassical Eilenberger equations in Refs. \onlinecite{SZ89,SZ}. Our
analysis (see also \cite{OGZB}) reproduces these results and generalizes
them to the case of nonuniform fluctuations of the field $V$.

Taking into account only uniform fluctuations of the
electric field one can also derive the Caldeira-Leggett
action expressed in terms of the {\it electron coordinate only}.
In the quasi-one-dimensional
situation one should simply write down the electron action on the Keldysh
contour, take into account the potential energy $-Ex$
of the electron with the coordinate $x(t)$ in the fluctuating
electric field $E(t)$ and add the
action for the field (\ref{actionf},\ref{eps})
(with the last term in (\ref{eps}) being dropped). After identification
$E(\omega ) =-ikV(\omega ,k) \to -V(\omega )/L$ ($L$ is the sample length)
and integration over the fluctuating field $V$ one arrives at
the Caldeira-Leggett action for the electron coordinate $x(t)$. In this
case the effective viscosity in the Caldeira-Leggett influence
functional is proportional to $1/\sigma$ (in contrast to the
effective viscosity for the field $V$ which is proportional to $\sigma$).
The whole procedure is completely analogous to that discussed in details
in Ref. \onlinecite{GZ92} where
we considered the real time effective action for a dissipative system
characterized by two collective degrees of freedom (the phase and
the charge). Integrating over the charge variable one arrives at the
Caldeira-Leggett action in the ``phase only'' representation. The same
can be done here if we use a formal analogy of $x(t)$ with the phase
and $V/i\omega$ with the charge (as defined in Ref. \onlinecite{GZ92}).

For our present purposes it is not sufficient to
restrict ourselves to uniform fluctuations of the collective
coordinate $V$ of the electron environment. We will see that
fluctuations with nonzero $k$ play an important role and should
be taken into account in the quantitative analysis. The corresponding
effective action will be derived in the next section. However the main
message is clear already from the simple example considered above:
in a disordered metal the effect of Coulomb interaction of the
electron with other electrons is equivalent to that of
an effective {\it dissipative} environment with the correspondent
effective viscosity governed by the Drude conductivity $\sigma$.

\section{Decoherence Time and Conductance}

\subsection{Conductance and electron effective action}

In order to evaluate the system conductance we will determine the
single particle density matrix $\bbox{\rho}$ in the presence
of an external electric potential $V_x(\bbox{r})$ applied to
the metal. Generalization of the results obtained in the
previous sections to the case $V_x\neq 0$ is straightforward.
The density matrix is determined by the equation
\begin{equation}
\bbox{\rho}(t)=\langle\rho_V(t,V_x))\rangle_{V_1,V_2}=\frac
{\int{\cal D}V_1{\cal D}V_2\; \rho_V(t,V_x)e^{i\bbox{S}_{V_x}[V_1,V_2]}}
{\int{\cal D}V_1{\cal D}V_2\; e^{i\bbox{S}_{V_x}[V_1,V_2]}};
\label{rho}
\end{equation}
where the effective action $\bbox{S}_{V_x}[V_1,V_2]$ reads:
\begin{eqnarray}
i\bbox{S}_{V_x}[V_1,V_2] &=&
2\text{Tr}\ln\hat G^{-1}_{V+V_x}+i\int\limits_0^t dt'
\int d\bbox{r}\frac{(\nabla V_1)^2-(\nabla V_2)^2}{8\pi},
\label{act11}
\end{eqnarray}
where the subscript $V+V_x$ indicates the shift of the fields
$V_{1,2}\rightarrow V_{1,2}+V_x$. Here we use the formally exact expression
for the effective action, the approximation (\ref{actionf}) will be introduced
after the expansion in $V_x$ will be carried out.

The density matrix $\rho_V(t,V_x)$ obeys the equation (\ref{rho5})
with $V^+ \to V^+ +V_x$. Assuming the field $V_x$ sufficiently small
one can linearize the equation for $\rho_V(t,V_x)=\rho_V(t)+
\delta\rho_V(t,V_x)$ and get
\begin{equation}
i\frac{\partial\delta\rho_V}{\partial t} =
H_1\delta\rho_V -
\delta\rho_V H_2 - [eV_x,\rho_V];
\label{deltarho}
\end{equation}
where
\begin{eqnarray}
H_1=H_0-eV^+ -\frac{1}{2}(1-2\rho_V)eV^-,
\nonumber \\
H_2=H_0-eV^+ +\frac{1}{2}eV^-(1-2\rho_V).
\label{Hj}
\end{eqnarray}

The formal solution of the equation (\ref{deltarho}) can be easily found:
\begin{equation}
\delta\rho_V(t)=
i\int\limits_0^t dt' U_1(t,t')[eV_x,\rho_V(t')] U_2(t',t),
\label{solution}
\end{equation}
where
\begin{equation}
U_{1,2}(t_1,t_2)={\rm\bf T}\exp\left[-i\int\limits_{t_1}^{t_2}
dt'\medskip H_{1,2}(t')\right].
\label{Uj}
\end{equation}
The operators $H_{1,2}$ (\ref{Hj}) are nonlocal (since they contain
the density matrix), therefore the path integral representation for the
evolution operators (\ref{Uj}) contains an additional integration over
momentum.  The operators (\ref{Hj}) can be written in the form:
\begin{eqnarray}
H_1(\bbox{p,r})=\frac{\bbox{p}^2}{2m}+U(\bbox{r}) - eV^+(t,\bbox{r})
-\frac{1}{2}\big[1-2n(H_0(\bbox{p},\bbox{r}))\big]eV^-(t,\bbox{r}),
\nonumber \\
H_2(\bbox{p,r})=\frac{\bbox{p}^2}{2m}+U(\bbox{r}) - eV^+(t,\bbox{r})
+\frac{1}{2}eV^-(t,\bbox{r})\big[1-2n(H_0(\bbox{p},\bbox{r}))\big],
\label{H12}
\end{eqnarray}
where $n(\xi)=1/[\exp(\xi/T)+1]$ is the Fermi function.
In deriving (\ref{H12}) from (\ref{Hj}) we set $\rho_V$ to be an
equilibrium density matrix. We also neglected the effect of
Coulomb interaction in the expression for $\rho_V(t)$.
This approximation is justified as long as Coulomb interaction is
sufficiently weak. Note that the same approximation for $\rho_V(t)$
can be used in eq. (\ref{solution}).

The evolution operators (\ref{Uj}) acquire the form:
\begin{equation}
U_{1,2}(t_1,t_2;\bbox{r}_f,\bbox{r}_i)=
\int\limits_{\bbox{r}(t_1)=\bbox{r}_i}^{\bbox{r}(t_2)=
\bbox{r}_f}{\cal D}\bbox{r}(t')\int{\cal D}\bbox{p}(t')
\;
\exp\left[i\int\limits_{t_1}^{t_2} dt'\big( \bbox{p\dot r} -
H_{1,2}(\bbox{p},\bbox{r})\big)\right].
\label{path}
\end{equation}

For the sake of generality we note that in the presence of
interaction there exists an additional (linear in the field)
correction to the density matrix. In order to see that
let us expand the action (\ref{act11}) to the first order in
$V_x$:
\begin{equation}
i\delta\bbox{S}[V_x]= -2e\text{tr}\big((G_{11}-G_{22})V_x\big).
\label{dS}
\end{equation}
This correction to the action gives an additional contribution to the density
matrix (\ref{rho}). Expressing the functions $G_{11}$ and $G_{22}$ in terms
of the density matrix $\rho_V$ and the evolution operators $u_{1,2}$
and combining this correction to the density matrix with one defined
in (\ref{solution}) we find
\begin{equation}
\delta\bbox{\rho}(t)=\langle\delta\rho_V\rangle_{V^+,V^-}+
\langle\delta\rho_{\text{int}}\rangle_{V^+,V^-},
\label{drho}
\end{equation}
where $\delta\rho_V$ is given by (\ref{solution}) and
$\delta\rho_{\text{int}}$ has the form
\begin{equation}
\delta\rho_{\text{int}}=-2i\rho_V(t)\int\limits_0^t dt'
\text{tr}\big(u_1(t,t')[eV_x(t'),\rho_V(t')]u_2(t',t)\big).
\label{drhoint}
\end{equation}
In the limit of weak interaction between electrons
the averaging in (\ref{drho}) may be performed with the approximate
action $S[V_1,V_2]$ (\ref{actionf}).

It is easy to observe that the second term in (\ref{drho})
is small in the limit of weak interaction and vanishes
completely if interaction is neglected. The weak
localization correction is described by the first term
in eq. (\ref{drho}) which will  be only considered further below.

Making use of a standard definition of the current density $\bbox{j}$:
\begin{equation}
\bbox{j}(t,\bbox{r})=\frac{ie}{m}
\left.\left(\nabla_{r_1}\delta\bbox{\rho}(t,\bbox{r}_1,\bbox{r}_2)
-\nabla_{r_2}\delta\bbox{\rho}(t,\bbox{r}_1,\bbox{r}_2)\right)
\right|_{\bbox{r}_1=\bbox{r}_2=\bbox{r}},
\label{j}
\end{equation}
combining it with the above equations and assuming the external electric
field to be constant in space and time, $V_x=-\bbox{Er}$, we arrive at the
expression for the system conductance
\begin{equation}
\sigma=
\frac{e^2}{3m}
\int\limits_{-\infty}^t dt'\int d\bbox{r}_{i1}d\bbox{r}_{i2}
\left.\left(\nabla_{r_{1f}}-\nabla{r_{2f}}\right)\right|_{\bbox{r}_{1f}
=\bbox{r_{2f}}}J(t,t';\bbox{r}_{1f},\bbox{r}_{2f};\bbox{r}_{1i},\bbox{r}_{2i})
(\bbox{r}_{1i}-\bbox{r}_{2i})\rho_0(\bbox{r}_{1i},\bbox{r}_{2i}).
\label{sigma}
\end{equation}
Here we have shifted the initial time to $-\infty$.
The function $J$ is the kernel of the operator
$$
\bbox{J}= \sum_V U|V\rangle\langle V|U^+
$$
where the sum runs over all possible states of the
electromagnetic environment. This function can be expressed in
terms of the path integral
\begin{eqnarray}
J(t,t';\bbox{r}_{1f},\bbox{r}_{2f};\bbox{r}_{1i},\bbox{r}_{2i})&=&
\int\limits_{\bbox{r}_1(t')=\bbox{r}_{1i}}^{\bbox{r}_1(t)=\bbox{r}_{1f}}
{\cal D}\bbox{r}_1
\int\limits_{\bbox{r}_2(t')=\bbox{r}_{2i}}^{\bbox{r}_2(t)=\bbox{r}_{2f}}
{\cal D}\bbox{r}_2\int{\cal D}\bbox{p}_1{\cal D}\bbox{p}_2\times
\nonumber \\
&&
\left\langle e^{iS_0[\bbox{r}_1,\bbox{p}_1]-iS_0[\bbox{r}_2,\bbox{p}_2]+
i\int_{t'}^tdt''\int d\bbox{r}
\big(f^-V^+ + f^+V^-\big)}\right\rangle_{V^+,V^-}.
\label{J11}
\end{eqnarray}
Here the action $S_0[x,p]$ has the form
\begin{equation}
S_0[\bbox{r},\bbox{p}]=\int\limits_{t'}^t dt''
\bigg(\bbox{p\dot r} - \frac{\bbox{p}^2}{2m} - U(\bbox{r})\bigg),
\label{S_0}
\end{equation}
and the ``charge densities'' $f^-,f^+$ are defined by the equations:
\begin{eqnarray}
f^-(t,\bbox{r})&=&e\delta (\bbox{r}-\bbox{r}_1(t))-e\delta
(\bbox{r}-\bbox{r}_2(t)),
\nonumber \\
f^+(t,\bbox{r})&=&\frac{1}{2}
\bigg(e\big[1-2n(\bbox{p}_1(t),\bbox{r}_1(t))\big]\delta
(\bbox{r}-\bbox{r}_1(t))+
e\big[1-2n(\bbox{p}_2(t),\bbox{r}_2(t))\big]
\delta (\bbox{r}-\bbox{r}_2(t))\bigg).
\label{f}
\end{eqnarray}

Averaging over $V^+,V^-$ in eq. (\ref{J11}) amounts to calculating
Gaussian path integrals with the action (\ref{actionf})
and can be easily performed. We obtain
\begin{eqnarray}
J(t,t';\bbox{r}_{1f},\bbox{r}_{2f};\bbox{r}_{1i},\bbox{r}_{2i})&=&
\int\limits_{\bbox{r}_1(t')=\bbox{r}_{1i}}^{\bbox{r}_1(t)=\bbox{r}_{1f}}
{\cal D}\bbox{r}_1\int\limits_{\bbox{r}_2(t')=\bbox{r}_{2i}}^{\bbox{r}_2(t)
=\bbox{r}_{2f}}{\cal D}\bbox{r}_2\int{\cal D}\bbox{p}_1{\cal D}
\bbox{p}_2\times
\nonumber \\
&&
\times \exp\big\{iS_0[\bbox{r}_1,\bbox{p}_1]-iS_0[\bbox{r}_2,\bbox{p}_2]-
iS_R[\bbox{r}_1,\bbox{p}_1,\bbox{r}_2,\bbox{p}_2]-
S_I[\bbox{r}_1,\bbox{r}_2]\big\};
\label{J}
\end{eqnarray}
where
\begin{eqnarray}
S_R[\bbox{r}_1,\bbox{p}_1,\bbox{r}_2,\bbox{p}_2]&=&
\frac{e^2}{2}\int\limits_{t'}^t dt_1 \int\limits_{t'}^t dt_2
\big\{R(t_1-t_2,\bbox{r}_1(t_1)-\bbox{r}_1(t_2))
\big[1-2n\big(\bbox{p}_1(t_2),\bbox{r}_1(t_2)\big)\big]-
\nonumber \\
&&
-R(t_1-t_2,\bbox{r}_2(t_1)-\bbox{r}_2(t_2))
\big[1-2n\big(\bbox{p}_2(t_2),\bbox{r}_2(t_2)\big)\big]
\nonumber \\
&&
+R(t_1-t_2,\bbox{r}_1(t_1)-\bbox{r}_2(t_2))
\big[1-2n\big(\bbox{p}_2(t_2),\bbox{r}_2(t_2)\big)\big]-
\nonumber \\
&&
-R(t_1-t_2,\bbox{r}_2(t_1)-\bbox{r}_1(t_2))
\big[1-2n\big(\bbox{p}_1(t_2),\bbox{r}_1(t_2)\big)\big]
\big\};
\label{SR}
\end{eqnarray}
and
\begin{eqnarray}
S_I[\bbox{r}_1,\bbox{r}_2]&=&
\frac{e^2}{2}\int\limits_{t'}^t dt_1 \int\limits_{t'}^t dt_2
\bigg\{I(t_1-t_2,\bbox{r}_1(t_1)-\bbox{r}_1(t_2))+
I(t_1-t_2,\bbox{r}_2(t_1)-\bbox{r}_2(t_2))-
\nonumber \\
&&
-I(t_1-t_2,\bbox{r}_1(t_1)-\bbox{r}_2(t_2))-
I(t_1-t_2,\bbox{r}_2(t_1)-\bbox{r}_1(t_2))\bigg\}.
\label{SI}
\end{eqnarray}
At the scales $|\bbox{r}|\gtrsim l$
the functions $R$ and $I$ are defined by the equations
\begin{eqnarray}
R(t,\bbox{r})&=&\int\frac{d\omega d^3k}{(2\pi)^4}\medskip
\frac{4\pi}{k^2\epsilon(\omega,k)}e^{-i\omega t+i\bbox{kr}}
\label{R}\\
I(t,\bbox{r})&=&\int\frac{d\omega d^3k}{(2\pi)^4}\medskip
{\rm Im}\left(\frac{-4\pi}{k^2\epsilon(\omega,k)}\right)
\coth\bigg(\frac{\omega}{2T}\bigg)
e^{-i\omega t+i\bbox{kr}} .
\label{RI}
\end{eqnarray}
If necessary, more general expressions for $R$ and $I$ for the
whole range $|\bbox{r}|\gtrsim 1/p_F$ can be easily derived.
We will avoid doing this here because the approximation (\ref{RI})
is already sufficient for our present purposes.

Note that the expression in the exponent of eq. (\ref{J}) defines
the real time effective action of the electron propagating in a
disordered metal and interacting with other electrons. The first
two terms represent the electron action $S_0$ (\ref{S_0}) on
two branches of the Keldysh contour
while the last two terms $S_R$ and $S_I$ determine the influence
functional which comes from the effective electron (and/or phonon)
environment. As can be seen from eqs. (\ref{SR}-\ref{RI}) this influence
functional is in general not identical to one derived in the
Caldeira-Leggett model. However on a qualitative level the similarity
is obvious: in both models the influence functionals describe the effect of a
certain effective {\it dissipative} environment.

Let us now analyze the expression (\ref{sigma})
and establish connection with other conductivity calculations.
It is convenient to introduce the Wigner function
$n(\bbox{p},\bbox{r})=\int d\bbox{r}^- e^{-i\bbox{pr}^-}
\rho(\bbox{r}+\bbox{r}^-/2,\bbox{r}-\bbox{r}^-/2)$
instead of the density matrix.
For homogeneous systems it does not depend on $\bbox{r}$ at the scales
exceeding the mean free path. With this in mind we obtain
\begin{equation}
\sigma
=-\frac{2e^2}{3m}\int\limits_{-\infty }^tdt^{\prime }\int \frac{d^3 {\bf
p}}{(2\pi )^3}{\bf p}\hat W(t,t^{\prime })\frac{\partial n({\bf p})}{\partial
{\bf p}};
\label{sigma3}
\end{equation}
where $\hat W (t,t^{\prime })$ is the evolution
operator for the Wigner function:
\begin{equation}
n(t,x,{\bf p})=\hat
W(t,t^{\prime })n(t^{\prime },x,{\bf p}).
\end{equation}
The kernel of this operator and that of $J$ (\ref{J}) are related
by means of the Fourier transformation with respect to
$\bbox{r}_{1i,1f}-\bbox{r}_{2i,2f}$.
In equilibrium one
has $n({\bf p})=1/(\exp (\xi /T)+1)$, and therefore at small $T$ one finds
\[
\frac{\partial n({\bf
p})}{\partial {\bf p}}\simeq -{\bf v}_F\delta (\xi )
\]
and arrives at the standard result (cf. e.g. \cite{CS})
\begin{equation}
\sigma
=\frac{2e^2N_0}3\int\limits_{-\infty }^tdt^{\prime }\langle {\bf v}(t) {\bf
v}(t^{\prime })\rangle .
\label{Schmid}
\end{equation}

\subsection{Weak localization correction and decoherence time}

Let us analyse the structure of the function $J$ (\ref{J}) in the same
spirit as it has been done in Ref. \onlinecite{CS}. In the
zero order approximation
one can neglect the terms $S_R$ and $S_I$ describing the effect
of Coulomb interaction. Then in the quasiclassical limit $p_Fl \gg 1$
the path integral (\ref{J}) is dominated by the saddle point trajectories for
the action $S_0$ which are just classical paths determined by the
Hamilton equations
\begin{equation}
\bbox{\dot p}=-\frac{\partial H_0(\bbox{p,r})}{\partial \bbox{r}},
\quad \bbox{\dot r}=\frac{\partial H_0(\bbox{p,r})}{\partial \bbox{p}}
\label{Hamilton}
\end{equation}
with obvious boundary conditions $\bbox{r}_1(t')=\bbox{r}_{1i}$,
$\bbox{r}(t)=\bbox{r}_{1f}$ for the action $S_0[\bbox{r}_1,\bbox{p}_1]$ and
$\bbox{r}_2(t')=\bbox{r}_{2i}$,
$\bbox{r}_2(t)=\bbox{r}_{2f}$ for the action $S_0[\bbox{r}_2,\bbox{p}_2]$.
Substituting these saddle point trajectories into (\ref{J}) and integrating
out small fluctuations around them one finds
\begin{eqnarray}
J(t,t';\bbox{r}_{1f},\bbox{r}_{2f};\bbox{r}_{1i},\bbox{r}_{2i})&=&
\sum\limits_{\bbox{r}_1}A_{\bbox{r}_1}\sum\limits_{\bbox{r}_2}A^*_{\bbox{r}_2}
\exp\big(iS_0(t,t';\bbox{r}_{1f},\bbox{r}_{1i})-
iS_0(t,t';\bbox{r}_{2f},\bbox{r}_{2i})\big)
\label{classics}
\end{eqnarray}
where the actions $S_0(t,t';\bbox{r}_{1,2f},\bbox{r}_{1,2i})$ are taken on
the classical pathes $\bbox{r}_{1,2}(t)$ and
\begin{eqnarray}
A_{\bbox{r}_1}=\sqrt{\frac{i^3}{8\pi^3}
\left|
{\rm det}\left(\frac{\partial^2S_0(\bbox{r}_{1f},\bbox{r}_{1i})}
{\partial \bbox{r}_{1f}
\partial \bbox{r}_{1i}}\right)\right|}.
\label{preexp}
\end{eqnarray}
The value $A^*_{\bbox{r}_2}$ is defined analogously.

Since in a random potential $U(\bbox{r})$ there is in general no
correlation between different classical paths $\bbox{r}_1(t)$ and
$\bbox{r}_2(t)$ these pathes give no contribution to the double sum
(\ref{classics}): the difference of two actions in the exponent of
(\ref{classics}) may have an arbitrary value and the result
averages out after summation. Thus only the paths for which
$S_0[\bbox{r}_1,\bbox{p}_1]\simeq S_0[\bbox{r}_2,\bbox{p}_2]$ provide
a nonvanishing contribution to (\ref{classics}). Two different classes
of such paths can be distinguished (see e.g. \cite{CS}):

i) The two classical paths are almost the same:
$\bbox{r}_1(t'')\simeq \bbox{r}_2(t'')$, $\bbox{p}_1(t'')\simeq \bbox{p}_2(t'')$ (see Fig. 1a).
For such pairs we obviously have $\bbox{r}_{1i}\simeq \bbox{r}_{2i}$ and
$\bbox{r}_{1f}\simeq \bbox{r}_{2f}$. In other words, in the path integral
(\ref{J}) one integrates only over trajectories with
$|\bbox{r}_1(t'')-\bbox{r}_2(t'')|\lesssim 1/p_F$. Physically this corresponds
to the picture of electrons propagating as nearly classical particles
which can be described by the diagonal elements of the density matrix only.
In the diffusive limit these paths give rize to diffusons (see e.g.
\cite{AAK1}) and yield the standard Drude conductance.

ii) The pairs of time reversed paths. In this case
$\bbox{r}_{1i}\simeq \bbox{r}_{2f}$, $\bbox{r}_{1f}\simeq \bbox{r}_{2i}$
(Fig. 1b). In the path integral (\ref{J}) the trajectories $\bbox{r}_1$
and $\bbox{r}_2$ are related as $\bbox{r}_2(t'')\simeq \bbox{r}_1(t+t'-t'')$
and $\bbox{p}_2(t'')\simeq -\bbox{p}_1(t+t'-t'')$. In other words,
in (\ref{J}) one integrates over paths with
$|\bbox{r}_2(t'')-\bbox{r}_1(t+t'-t'')|\lesssim 1/p_F$, however
the difference $|\bbox{r}_1(t'')-\bbox{r}_2(t'')|$ may be
arbitrarily large in this case. In a disordered metal these paths
essentially determine the dynamics of off-diagonal elements of the
electron density matrix. They correspond to Cooperons and give rize
to the weak localization correction to conductivity.
This correction is defined by the following equation \cite{GLK,AAK1,CS}
\begin{equation}
\delta\sigma=-\frac{2e^2D}{\pi}\int\limits_{\tau_e}^{\infty}
dt W(t)=
-\frac{2e^2D}{\pi}\int\limits_{-\infty}^{t-\tau_e}dt' W(t-t').
\label{dsigma}
\end{equation}
Here we
changed the parameter of integration in order to make
the relation with the equation (\ref{sigma}) more transparent.
The quantity $W(t)$ represents the effective
probability for the diffusive path to return to the same
point after the time $t$. Note that $W(t)$ contains the contribution from
time reversed paths and therefore differs from a classical probability.
However, in the absence of any kind of interaction
which breaks the time reversal symmetry this value coincides with the classical
return probability and is given by the formula
$W_0(t)=(4\pi Dt)^{-d/2}a^{-(3-d)}$, where $d$ is the system dimension
and $a$ is the transverse sample size (the film thickness for $d=2$ and the
square root of the wire cross section for $d=1$).

The weak localization correction (\ref{dsigma}) diverges for $d \leq 2$.
This divergence can be cured by
introducing the upper limit cutoff for the integral (\ref{dsigma})
at a certain time $\tau_\varphi$. This time is usually
reffered to as decoherence time. As we have already discussed,
in a disordered metal the time $\tau_{\varphi}$ is determined by
electron-electron, electron-phonon and other types of interaction which
may destroy quantum coherence. From (\ref{dsigma}) one finds \cite{AAK1,CS}:
\begin{equation}
\delta\sigma_d=\left\{\begin{array}{ll}
           -\frac{\sqrt{3}e^2}{2\pi^{3/2}l}, & d=3, \\
    -\frac{e^2}{2\pi^2 }\ln\left(\frac{\tau_\varphi}{\tau_e}\right), & d=2 \\
    -\frac{e^2}{\pi }\sqrt{D\tau_\varphi}, & d=1.
     \end{array}\right.
\label{dsigma1}
\end{equation}
Here and below $\sigma_d = \sigma a^{3-d}$ is the Drude conductance of a
$d$-dimensional sample.

\begin{figure}[h]
\centerline{\psfig{file=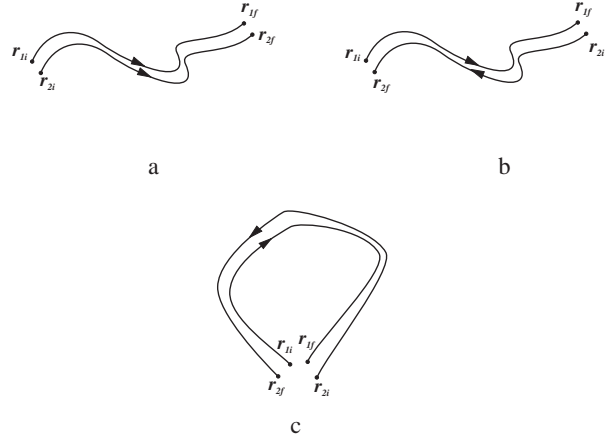,height=6cm}}
\caption{Different quasiclassical trajectories: a) $\bbox{r}_1$
and $\bbox{r}_2$ are close to each other; b) time reversed paths;
c) time reversed paths returning to
the vicinity of the initial point, these paths determine the weak
localization correction to conductivity}
\label{paths}
\end{figure}

To evaluate $\tau_\varphi$ we first note that the functions $R$ and $I$
(\ref{RI}) change slowly at distances of the order of the Fermi
wavelength $1/p_F$. Therefore we may put $\bbox{r}_1(t'')=\bbox{r}(t'')$,
$\bbox{r}_2(t'')=\bbox{r}_t(t'')\equiv\bbox{r}(t+t'-t'')$.
Here $\bbox{r}(t'')$ is a classical trajectory with the initial point
$\bbox{r}(t')=0$ (its position can be chosen arbitrarily) and the final
point $|\bbox{r}(t)|\lesssim l$ (Fig. 1c). In other words, we
consider trajectories
which return to the vicinity of the initial point.
Then the return probability $W(t)$ can be written in the form
\begin{equation}
W(t-t')\simeq W_0(t-t')
\left\langle e^{-iS_R[t,t';\bbox{r},\bbox{p};\bbox{r}_t,\bbox{p}_t]
-S_I[t,t';\bbox{r},\bbox{r}_t]}
\right\rangle_{r},
\label{W}
\end{equation}
where the average is taken over all diffusive paths returning to the initial
point.
The average in (\ref{W}) decays exponentially in time, therefore we
may define $\tau_\varphi$ as follows:
\begin{eqnarray}
e^{-\frac{(t-t')}{\tau_\varphi}}&=&
\left\langle e^{-iS_R[t,t';\bbox{r},\bbox{p};\bbox{r}_t,\bbox{p}_t]
-S_I[t,t';\bbox{r},\bbox{r}_t]}
\right\rangle_{r}
\nonumber \\
&\simeq&
e^{\left\langle -iS_R[t,t';\bbox{r},\bbox{p};\bbox{r}_t,\bbox{p}_t]
-S_I[t,t';\bbox{r},\bbox{r}_t]
\right\rangle_{r}}
,\qquad (t-t')\rightarrow +\infty;
\label{tau11}
\end{eqnarray}
where the average is again taken over all classical paths returning
to the same point at the time $t$.

Let us consider the term $S_R$ (\ref{SR}). Making use of
the obvious relations $R(t,-\bbox{r})=R(t,\bbox{r})$,
$n(-\bbox{p},\bbox{r})=n(\bbox{p},\bbox{r})$
after a trivial algebra we get:
\begin{eqnarray}
S_R[t,t';\bbox{r,p;r}_t,\bbox{p}_t]&=&
\frac{e^2}{2}\int\limits_{t'}^t dt_1 \int\limits_{t'}^t dt_2
\bigg\{2R(t_1-t_2,\bbox{r}(t_1)-\bbox{r}(t_2))
\big[n(\bbox{p}(t_1),\bbox{r}(t_1))-n(\bbox{p}(t_2),\bbox{r}(t_2))\big]+
\nonumber \\
&&
+R(t_1+t_2-t-t', \bbox{r}(t_1)-\bbox{r}(t_2))
\big[1-2n(\bbox{p}(t_2),\bbox{r}(t_2))\big]-
\nonumber \\
&&
-R(t_1+t_2-t-t', \bbox{r}(t+t'-t_1)-\bbox{r}(t+t'-t_2))
\big[1-2n(\bbox{p}(t+t'-t_2),\bbox{r}(t+t'-t_2))\big]\bigg\}
\label{SR1}
\end{eqnarray}

It is clear that the difference of the occupation numbers $n$ in the first term
of eq. (\ref{SR1}) vanish: $n$ depends only on the energy and not on time
because the energy is conserved along the classical path (\ref{Hamilton}).
Thus the first term in the integral (\ref{SR1}) is identically zero already
before averaging over diffusive paths. The difference of two other terms
vanishes after averaging over such paths. It can be easily seen
if we again bear in mind that the occupation numbers do not depend on
time and make use of the fact that the average
$\langle R(t_3,\bbox{r}(t_1)-\bbox{r}(t_2))\rangle_r$ is a function of
the time $t_3$ and the absolute value of the time difference $|t_1-t_2|$.
This implies that after averaging the two last terms in eq. (\ref{SR1})
are equal and cancel each other exactly. As a result the whole functional
$S_R$ does not give any contribution to $1/\tau_\varphi$.

The same analysic can be applied to $S_I$. We find
\begin{eqnarray}
S_I[t,t';\bbox{r;r}_t]&=&
\frac{e^2}{2}\int\limits_{t'}^t dt_1 \int\limits_{t'}^t dt_2
\bigg\{2I(t_1-t_2,\bbox{r}(t_1)-\bbox{r}(t_2))-
\nonumber \\
&&
-I(t_1+t_2-t-t', \bbox{r}(t_1)-\bbox{r}(t_2))-
I(t_1+t_2-t-t', \bbox{r}(t+t'-t_1)-\bbox{r}(t+t'-t_2))
\bigg\}
\label{SI1}
\end{eqnarray}
Averaging over diffusive paths $\bbox{r}(t)$ and taking the limit
$t-t'\rightarrow +\infty$, we observe that $\tau_\varphi$ is determined by the
first term under the integral, the other terms give only the time independent
(and therefore irrelevant) constant.
Thus we get
\begin{equation}
\frac{1}{\tau_\varphi}=
e^2\int\limits_{-\infty}^{+\infty} dt
\langle I(t,\bbox{r}(t)-\bbox{r}(0))\rangle_{r}
\label{tau}
\end{equation}
To find the average over the diffusive paths, we introduce the Fourier
transform of the function $I(t,\bbox{r})$ and replace
$\langle e^{-ik(\bbox{r}(t)-\bbox{r}(t'))}\rangle_{r}$ by $e^{-Dk^2|t-t'|}$.
Then we obtain
\begin{equation}
\frac{1}{\tau_\varphi}=
\frac{e^2}{a^{3-d}}\int\limits_{-\infty}^{+\infty}dt
\int\frac{d\omega d^dk}{(2\pi)^{d+1}} {\rm Im}
\left(\frac{-4\pi}{k^2\epsilon(\omega,k)}\right)
\coth\left(\frac{\omega}{2T}\right)
e^{-i\omega t-Dk^2|t|}
\label{3d}
\end{equation}
As it was already discussed, eq. (\ref{3d}) includes the effect of both
electron-electron and electron-phonon interactions.

To evaluate $\tau_\varphi$ we use the expression (\ref{eps}) for the dielectric
susceptibility. For typical metallic systems one can usually neglect
the first term in the expression for $\epsilon(\omega,k)$ (\ref{eps}).
Then we find
\begin{equation}
{\rm
Im}\left(\frac{-1}{\epsilon(\omega,k)}\right)=
\frac{\omega}{4\pi\sigma}\left[1+
\frac{c^4k^4-c^4\omega^2/D^2}
{(\omega^2-c^2k^2)^2+c^4\omega^2/D^2}\right].
\label{im}
\end{equation}
Here $c=\omega_{pi}(D/4\pi\sigma )^{1/2}$ is the speed of sound in our
system. Possible corrections to (\ref{im}) may become important only
in special cases of 1d and 2d semiconductor systems, where the capacitance may
become important.  We will come back to this point later.

The expression (\ref{im}) can be significantly simplified for
$c/D\lesssim k\lesssim 1/l$. In this limit we obtain
\begin{equation}
{\rm Im}\left(\frac{-1}{\epsilon(\omega,k)}\right)\simeq
{\rm Im}\left(\frac{-1}{\epsilon(\omega,k)}\right)_{ee}+
{\rm Im}\left(\frac{-1}{\epsilon(\omega,k)}\right)_{eph},
\label{imeps}
\end{equation}
where
\begin{equation}
{\rm Im}\left(\frac{-1}{\epsilon(\omega,k)}\right)_{ee}=
\frac{\omega}{4\pi\sigma},
\label{imee}
\end{equation}
and
\begin{equation}
{\rm Im}\left(\frac{-1}{\epsilon(\omega,k)}\right)_{eph}=
\frac{Dck^3}{8\sigma}\big(\delta(\omega-ck)-\delta(\omega+ck)\big).
\label{imeph}
\end{equation}
Phonons with small wave vectors $k\lesssim c/D$ are strongly damped.
For such $k$ we may put
${\rm Im}\left(\frac{-1}{\epsilon(\omega,k)}\right)_{eph}\simeq 0$.

\subsection{Results}

With the aid of the above results we can now calculate the decoherence
time $\tau_{\varphi}$. Let us first take into account only
electron-electron contribution to $\epsilon$ and obtain the result for
a quasi-one-dimensional system with $a \lesssim l$.
Substituting (\ref{imee}) into (\ref{3d})
and integrating over time and the wave vector we arrive at the
integral over $\omega$ which diverges at both low and high frequencies.
The low frequency divergence is cured in a standard manner \cite{FN0} by
neglecting the effect of environmental fluctuations with frequencies below
$1/\tau_{\varphi}$.
At high frequencies the integral should be cut at the scale of the order
of the inverse transport time because at higher $\omega$ the approximation
of electron diffusion becomes incorrect. Then we obtain
\begin{equation}
\frac{1}{\tau_{\varphi}(T)}=\frac{e^2\sqrt{2D}}{\sigma_1 }
\int\limits_{1/\tau_{\varphi}}^{1/\tau_{e}}
\frac{d\omega}{2\pi}\frac{\coth (\omega /2T)}{\sqrt{\omega}}.
\label{tau1}
\end{equation}
Eq. (\ref{tau1}) yields
\begin{equation}
\frac{1}{\tau_{\varphi}}=
\frac{e^2}{\pi\sigma_1 }\sqrt{\frac{2D}{\tau_e}}
\left[2T\sqrt{\tau_e\tau_\varphi} + 1\right].
\label{tau2}
\end{equation}
At sufficiently high
temperature the first term dominates and the standard result \cite{AAK}
$\tau_{\varphi} \sim (\sigma_1 /e^2D^{1/2}T)^{2/3}$ is recovered. This is
a classical contribution to $\tau_{\varphi}$. As $T$ is
lowered the number of classical (low frequency) modes decreases
and eventually vanishes in the limit $T \to 0$. At
$T \lesssim T^{(1)}_q \sim 1/\sqrt{\tau_{\varphi}\tau_e}$ the expression (\ref{tau2})
is dominated by the second term and $\tau_{\varphi}$ saturates at the value
\begin{equation}
\tau_{\varphi} \approx \pi \sigma_1 /e^2v_F
\label{tau4}
\end{equation}
The estimate for the crossover temperature $T^{(1)}_q$ is obvious from
(\ref{tau2},\ref{tau4}):
\begin{equation}
T^{(1)}_q \approx ev_F/2\sqrt{\sigma_1 l}.
\label{Tq}
\end{equation}
Making use of eqs. (\ref{dsigma1},\ref{tau4}) it is also easy to find the weak localization
correction $\delta \sigma_1$ to the Drude conductance in the limit $T=0$.
For $T \lesssim T^{(1)}_q$ we obtain
\begin{equation}
\frac{\delta \sigma_1}{\sigma_1}=-\frac{e^2}{\pi \sigma_1
}\sqrt{D\tau_{\varphi}} \approx - \frac1{p_Fs^{1/2}}, \label{delsig}
\end{equation}
i.e. $\delta \sigma_1 \approx - \sigma_1 /\sqrt{N_{ch}}$, where $N_{ch} \sim p_F^2s$ is
the effective number of conducting channels in a 1d mesoscopic system.

For 2d and 3d systems the same analysis yields
\begin{eqnarray}
\frac{1}{\tau_\varphi} & = & \frac{e^2}{4\pi\sigma_2 \tau_e}
[1+2T\tau_e\ln(T\tau_\varphi)],  \; \; \; \; \; \; \;  \quad{\rm 2d},
\nonumber \\
\frac{1}{\tau_\varphi} & = & \frac{e^2}
{3\pi^2\sigma\sqrt{2D}\tau_e^{3/2}}[1+6(T\tau_e)^{3/2}],\;  \quad{\rm 3d},
\label{11}
\end{eqnarray}

According to (\ref{11}) in 2d and 3d systems the decoherence time
becomes independent of $T$ already at relatively high temperatures.
In the 3d case such temperatures can be of the order of the inverse
transport time, while for 2d systems we have $T^{(2)}_q \sim
(2\tau_e\ln (p_F^2al))^{-1}$.

Note that in the 2d case we again assumed $a \lesssim l$. Only
provided this condition is satisfied the above results for 1d and 2d
systems ave valid for the whole temperature range. At sufficiently
high temperatures this condition can be softened because in this
case $\tau_{\varphi}$ is determined by the low frequency fluctuations
of the environment.  Then the system can be considered as a quasi-1d
(quasi-2d) one if its transversal dimension is smaller that the
corresponding phase breaking length $a \ll L_{\varphi} \sim
\sqrt{D\tau_{\varphi}}$. However
at low temperatures high frequency modes become important and the
situation changes. E.g. in the limit $a \gg l$ the diffusion
process has two (and sometimes even three) stages: at short times
(i.e. at frequencies higher than $D/a^2$) diffusion is obviously 3d,
whereas for longer times it can be 1d or 2d.

For $a \gg L_{\varphi}$ the system is obviously 3d at all $T$.
In the intermediate case $l \ll a \ll L_{\varphi}$ one should use
the corresponding 1d or 2d formulas for $\tau_{\varphi}$ at high temperatures
$T \gg T^{(d)}_{qa}$ and the 3d result (\ref{11}) in the low temperature
limit $T \ll T^{(d)}_{qa}$. The crossover temperature
$T^{(d)}_{qa}$ can be determined either directly from the integral (\ref{3d})
or just by comparison of the corresponding expressions for $\tau_{\varphi}$.
We get $T^{(d)}_{qa} \approx T^{(d)}_{q}(a/\pi l)^{3-d}$. Thus
in the limit $a \gg l$ the low temperature saturation of $\tau_{\varphi}$
takes place already at somewhat higher temperatures than in the case of
small $a$. The saturation value of $\tau_{\varphi}$ becomes
smaller for large $a$ since it is defined by the 3d result (\ref{11}).
Such details may be quite important for a quantitative comparison
with experimental results (see below).

The above discussion applies to metallic systems. In the case
of two-dimensional electron gas in semiconductors with
$l \ll a \ll L_{\varphi}$ the same arguments hold, but the saturation
value $\tau_{\varphi}$ at low $T$ is given by 2d (not 3d) result.
The classical-to-quantum crossover takes place at $T \sim T^{(1)}_q(a/4\pi l)$.

Now let us analyze the effect of the electron-phonon interaction
on the decoherence time $\tau_{\varphi}$. Substituting (\ref{imeph}) into
(\ref{3d}) and applying the same cutoff procedure after a simple integration
one obtains
$$
\frac{1}{\tau_\varphi^{eph}}=
%\left\{\begin{array}{ll}
\frac{e^2c}{4\pi^2\sigma_1}\ln\left(\frac{v_F}{3c}\right)+
\frac{e^2D}{4\pi\sigma_1c}T,\;\; \;\;\;\;\;\;\;\;\;\;\;\;\;\;\;\; \quad{\rm 1d},
$$
\begin{equation}
\frac{1}{\tau_\varphi^{eph}}=
\frac{e^2c}{8\pi^2\sigma_2v_F\tau_e}+
\frac{e^2}{4\pi^2\sigma_2} T
\ln\left(\min\left\{\frac{v_F}{3c},\frac{DT}{c^2}\right\}\right), \;\;
\quad{\rm 2d},
\label{phonons}
\end{equation}
$$
\frac{1}{\tau_\varphi^{eph}}=
\frac{e^2c}{16\pi^3\sigma v_F^2\tau_e^2}+
\frac{e^2}{4\pi^3\sigma v_F\tau_e}T
\min\left\{1,\frac{Tl}{c} \right\},\;\;\; \quad{\rm 3d}.
$$
%\end{array}\right.
%\label{phonons}
%\end{equation}
Comparing these expressions with the above results for $\tau_{\varphi}$
we find $\tau_\varphi^{ee}/\tau_\varphi^{eph} \sim c/v_F \ll 1$, i.e.
at low $T$ the decoherence effect due to zero-point fluctuations
of lattice ions is always much smaller than that of
quantum fluctuations in the electron bath.
The temperature dependent term in the expression for
the inverse dephasing time due to phonons is of the same order as
the corresponding electro-electron term in 1d and 2d, and can be even bigger
in 3d at relatively high temperatures $T\gtrsim c/l$. Thus we can
conclude that in 1d and 2d systems the decoherence effect due to
phonons is never important, while in 3d systems it can modify
the temperature dependence of the decoherence time at sufficiently high $T$.

The above results in 1d and 2d are valid if the number of conducting
channels $N_{ch}$ in the system is sufficiently big. In typical metallic
systems this condition is usually well satisfied. However, in semiconductors
one can in principle achieve the situation with $N_{ch} \lesssim 10$. In this
case the first term in the expression for $\epsilon$ (\ref{eps}) cannot
be neglected in general. Moreover for 1d and 2d samples with small
$N_{ch}$ the energy of the electromagnetic field outside the sample
may also give a substantial contribution.
In order to account for thhis effect we introduce the effective capacitance of
the system $C$. Then the influence functional for the field $V$ has the
form (see also \cite{OGZB})
\begin{eqnarray}
iS_C&=&
i\int\frac{d\omega d^dk}{(2\pi)^{d+1}}
V^-(-\omega,-k)\bigg[\frac{C(\omega,k)}{2}+
\frac{k^2
\big(\epsilon(\omega,k)-1\big)}{4\pi}\bigg]V^+(\omega,k)-
\nonumber \\
&&
-\frac{1}{2}
\int\frac{d\omega d^3k}{(2\pi)^4}
V^-(-\omega,-k)\frac{k^2{\rm Im}\epsilon(\omega,k)}{4\pi}
\coth\left(\frac{\omega}{2T}\right)
V^-(\omega,k);
\label{actionC}
\end{eqnarray}
where $C(\omega,k)\simeq (1+\epsilon_{s}(\omega))/4\ln(1/ka)$ for a 1d wire and
$C(\omega,k)=(1+\epsilon_s(\omega))k/8\pi$ for a 2d film. Here
$\epsilon_s(\omega)$ is the dielectric susceptibility of the substrate.
The Fourier transform of the function $I(t,\bbox{r})$ takes the form
\begin{equation}
I_{\omega,k}=
\langle|V^+_{k,\omega}|^2\rangle=
\frac{\omega\coth\left(\frac{\omega}{2T}\right)}
{\frac{(\omega C(\omega,k))^2}{\sigma_d k^2}+
\sigma_d q^2(1+\frac{C(\omega,k)D}{\sigma_d})^2}.
\label{VV}
\end{equation}
Substituting this expression into (\ref{tau}) we get for 1d wire
\begin{equation}
\frac{1}{\tau_{\varphi}(T)}=\frac{e^2\sqrt{2D}}{\sigma_1 }
\int\limits_{1/\tau_{\varphi}}^{1/\tau_{e}}
\frac{d\omega}{2\pi}\frac{\coth (\omega /2T)}{\sqrt{\omega}(1+2f(\omega))}
\bigg(1-\sqrt{\frac{f(\omega)}{1+f(\omega)}}\bigg),
\label{tauC}
\end{equation}
where $f(\omega)=\frac{(1+\epsilon_{s}(\omega))D}{4\sigma_1\ln(1/k_0a)}$ and
the value $k_0a$ is roughly of order one. We estimate
\begin{equation}
f(\omega)\sim \frac{1+\epsilon_s(\omega)}{e^2N_0s}
\sim \frac{\big(1+\epsilon_s(\omega)\big)p_Fr_B}{N_{ch}},
\label{ff}
\end{equation}
where $r_B=1/me^2\simeq 0.5$\AA$\;$ is the Bohr radius. For metallic wires
$p_Fr_B\sim 1$, $N_{ch}\gg 1$ and the function $f(\omega)$ is usually small
unless $\epsilon_s(\omega)\gg 1$ at frequencies of the order of
$1/\tau_e$. However for semiconductors $f(\omega)$ may be large and
$\tau_\varphi$ may become significantly longer than one could expect from
eq. (\ref{tau2}).

The same analysis can be carried out for 2d films. In this case the
effect of capacitance is described by the function
$$
f_2(\omega)=\frac{(1+\epsilon_s(\omega))\sqrt{D\omega}}{8\pi\sigma_2}
\sim
\frac{(1+\epsilon_s(\omega))\sqrt{\omega\tau_e}r_B}{p_Fla}.
$$
Again one can conclude that this effect is typically negligible for metallic
films. For semiconductors with small $N_{ch}$ the above effect
might cause an increase of $\tau_\varphi$.

\section{Quantum kinetic approach and Langevin equation}

Let us now demonstrate how the usual quantum kinetic description can be
derived from our analysis. We start from the equation for the density matrix
$\rho_V$ (\ref{rho5}). Rewriting this equation in the ``interaction
representation'', i.e. substituting $\rho_V\rightarrow e^{-iH_0t}\rho_V
e^{iH_0t}$ we find
\begin{equation}
i\frac{\partial\rho_V}{\partial t}=-e\hat V^+(t)\rho_V+\rho_V e\hat V^+(t)
-\frac{e}{2}\big((1-\rho_V)\hat V^-(t)\rho_V + \rho_V \hat V^-(t)(1-\rho_V)\big),
\label{rho2}
\end{equation}
where $\hat V^{\pm}(t)=e^{iH_0t}V^{\pm}(t)e^{-iH_0t}$.
Let us integrate this equation over time,
%\begin{equation}
%\rho(t)=ie\int\limits_{-\infty}^t dt'\bigg[\hat V^+(t')\rho(t')-\rho(t') \hat
%V^+(t') +\frac{1}{2}\big((1-\rho(t'))\hat V^-(t')\rho(t')+
%\rho(t')\hat V^-(t')(1-\rho(t'))\big)\bigg]
%\label{rho3}
%\end{equation}
then substitute the resulting expression for $\rho_V$ into the right hand side of
eq. (\ref{rho2}) and average over $V^{\pm}$.
If the Coulomb interaction is sufficiently weak one can proceed perturbatively
in $V$ and neglect the dependence of the density matrix $\rho_V$ on this
field in the right hand side of the resulting equation.
Then the result of averaging can be expressed in terms of the
correlation functions $\langle VV\rangle$. More precisely, two such functions
turn out to be important:
\begin{eqnarray}
\langle V^+(t_1,\bbox{r}_1)V^+(t_2,\bbox{r}_2) \rangle & = &
I(t_1-t_2,\bbox{r}_1-\bbox{r}_2),
%\qquad
\nonumber \\
\langle V^+(t_1,\bbox{r}_1)V^-(t_2,\bbox{r}_2)\rangle & = &
iR(t_1-t_2,\bbox{r}_1-\bbox{r}_2),
\label{VV1}
\end{eqnarray}
The function $\langle V^+(t_1,\bbox{r}_1)V^-(t_2,\bbox{r}_2)\rangle$
differs from zero only for $t_1>t_2$. The correlation function $
\langle V^-V^-\rangle $ is zero for all times. Taking this into account
we obtain
\begin{eqnarray}
\frac{\partial\bbox{\rho}}{\partial t}&=&
e^2\int\limits_{-\infty}^{t} dt'
\bigg\langle -\hat V^+(t)\hat V^+(t')\rho(t')+
\hat V^+(t)\rho(t')\hat V^+(t') +
\hat V^+(t')\rho (t')\hat V^+(t) -
\rho (t')\hat V^+(t')\hat V^+(t)-
\nonumber \\
&&
-\frac{1}{2}\hat V^+(t)(1-\rho (t'))\hat V^-(t')\rho (t')-
\frac{1}{2}\hat V^+(t)\rho (t')\hat V^-(t')(1-\rho (t'))+
\frac{1}{2}(1-\rho (t'))\hat V^-(t')\rho (t')V^+(t)+
\nonumber \\
&&
+\frac{1}{2}\rho (t')\hat V^-(t')(1-\rho (t'))V^+(t)\bigg\rangle_{V^+,V^-},
\label{eq}
\end{eqnarray}
where $\rho (t)=\rho_{V=0}(t)$.

For simplicity let us consider a clean metal. Making use of the momentum
conservation one can significanly simplify the equation (\ref{eq}).
In this case the density matrix is given by
$\rho(\bbox{r}_1-\bbox{r}_2)=
\int\frac{d^3p}{(2\pi)^3} n_p e^{i\bbox{p}(\bbox{r}_1-\bbox{r}_2)}$.
The operator $e^{-iH_0t}$ reduces to $e^{-i\xi_pt}$. Performing
the averaging with the aid of eqs. (\ref{VV1},\ref{R},\ref{RI}) we find
\begin{eqnarray}
\frac{dn_p}{dt}&=&
%\frac{e^2}{\pi^2}
%\int d\omega d^3k
%\delta(\omega+\xi_{p-k}-\xi_p)
%\bigg[\frac{\langle V^+V^+\rangle_{\omega,k}}{4\pi}(n_{p-k}-n_p)-
% {\rm Im}\left(\frac{-1}{k^2\epsilon(\omega,k)}\right)
%\big(n_p(1-n_{p-k})+n_{p-k}(1-n_p)\big)\bigg]=
%\nonumber \\
%&=&
\frac{e^2}{\pi^2}
\int d\omega d^3k {\rm Im}\left(\frac{-1}{k^2\epsilon(\omega,k)}\right)
\delta(\omega+\xi_{p-k}-\xi_p)
\bigg[\coth\left(\frac{\omega}{2T}\right)(n_{p-k}-n_p)-
n_p(1-n_{p-k})-n_{p-k}(1-n_p)\bigg].
\label{kinetic1}
\end{eqnarray}
The right hand side of this equation represents the standard collision integral
which vanishes in equilibrium, i.e. for $n_p=1/(\exp(\xi_p/T)+1)$.

The equation (\ref{kinetic1}) can be also rewritten in the following form:
\begin{eqnarray}
\frac{dn_p}{dt}&=&\frac{2e^2}{\pi^2}
\int\limits_{0}^{+\infty}
d\omega \int d^3k {\rm Im}\left(\frac{-1}{k^2\epsilon(\omega,k)}\right)
\bigg\{
\delta(\omega+\xi_{p-k}-\xi_p)N_{\omega}n_{p-k}(1-n_p)-
\nonumber \\
&&
-\delta(\omega+\xi_{p-k}-\xi_p)(1+N_{\omega})n_p(1-n_{p-k})
+\delta(\omega+\xi_p-\xi_{p-k})(1+N_{\omega})n_{p-k}(1-n_p)-
\nonumber \\
&&
-\delta(\omega+\xi_p-\xi_{p-k})N_{\omega}n_p(1-n_{p-k})\bigg\},
\label{kinetic}
\end{eqnarray}
where $N_\omega=1/(\exp(\omega/T)-1)$ is Bose function. This equation
describes the standard photon absorbtion and emission processes
and thus establishes a transparent relation between our approach and one
describing the kinetics of an electron interacting with the
quantized electromagnetic field. In our case the field $V$
is due to fluctuations of conducting electrons
(or lattice ions -- see below). It is quite clear however that
the physical nature of this field is not important for
the electron dynamics, at least as long as this Bose field remains
in equilibrium.

%have transparent interpretation: the first term describes the
%absorption of the energy by the electron with the momentum $p-k$ with the final
%momentum after absorption being $p$, the second term corresponds to the
%emission of the energy quantum $\omega$ by the electron with the momentum $p$,
%the third term describes the emission by the electron with the momentum $p-k$
%and, finally, the fourth term corresponds to the energy absorption by the
%electron with the momentum $p$. These four terms have the standard form, which
%is typical for the interaction of the electrons with some Bose field.

It is important to emphasize that the effect of electron-phonon interaction
is also taken into account in eq. (\ref{kinetic}). The phonon spectrum is
determined by the equation $\epsilon(\omega,k)=0$, i.e. the function
$\frac{-1}{\epsilon(\omega,k)}$ has a pole at $\omega=\omega_{ph}(k)-i0$.
Therefore one can write
\begin{eqnarray}
\frac{-1}{\epsilon(\omega,k)}=-\frac{A(k)}{\pi(\omega-\omega_{ph}(k)+i0)}+...
\nonumber \\
{\rm Im}\left(\frac{-1}{\epsilon(\omega,k)}\right)=
A(k)\delta(\omega-\omega_{ph}(k))+...,
\label{ph}
\end{eqnarray}
where other contributions to $\epsilon^{-1}$ are denoted by dots.
The value $A(k)$ determines the strength of electron-phonon
interaction. Within the simple model (\ref{imeph}) one has
$A(k)=Dck^3/8\sigma=ck^3/16e^2N_0$.

Substituting the expression (\ref{ph}) into (\ref{kinetic}) and
integrating over $\omega$ we reproduce the standard
electron-phonon collision integral:
\begin{eqnarray}
I_{eph}&=&\frac{2e^2}{\pi^2}
\int d^3k\medskip \frac{A(k)}{k^2}
\bigg\{
\delta(\omega_{ph}(k)+\xi_{p-k}-\xi_p)N_{\omega_{ph}(k)}n_{p-k}(1-n_p)-
\nonumber \\
&&
-\delta(\omega_{ph}(k)+\xi_{p-k}-\xi_p)(1+N_{\omega_{ph}(k)})n_p(1-n_{p-k})
+\delta(\omega_{ph}(k)+\xi_p-\xi_{p-k})(1+N_{\omega_{ph}(k)})n_{p-k}(1-n_p)-
\nonumber \\
&&
-\delta(\omega_{ph}(k)+\xi_p-\xi_{p-k})N_{\omega_{ph}(k)}n_p(1-n_{p-k})\bigg\},
\label{eph1}
\end{eqnarray}
This result demostrates that the function $\epsilon(\omega,k)$
correctly describes both electron-phonon and
electron-electron interactions. It is not surprizing, because
this function just accounts for the collective effect of the
environment. Electrons propagating in a metal
``feel'' only the fluctuating electric field produced by the environment,
both by electrons and lattice ions. Therefore it is quite natural that
both contributions can be successfully treated within the same approach.

The equations (\ref{kinetic1}-\ref{eph1}) are applicable if the distribution
functions $n_p$ and $N_\omega$ are close to the equilibrium Fermi and Bose
functions. It is not difficult to generalize this approach for stronger
deviations from equilibrium. Actually the electron-phonon collision
integral (\ref{eph1}) remains the same in this case, only the
distribution function $N_k$ can deviate far from the Bose function.
In order to generalize the electron-electron collision integral
we make use of the following nonequilibrium formulas:
\begin{eqnarray}
{\rm Im}\left(\frac{-1}{\epsilon(\omega,k)}\right)&=& -\frac{e^2}{\pi
k^2|\epsilon(\omega,k)|^2}\int d^3p
\delta(\omega-\xi_{p+k}+\xi_p)\big(n_{p+k}-n_p\big),
\nonumber \\
\langle V^+V^+ \rangle_{\omega,k}&=&
\frac{4e^2}{k^4|\epsilon(\omega,k)|^2}
\int d^3p \big[ n_{p+k}(1-n_p) + n_p(1-n_{p+k})\big]
\delta(\omega-\xi_{p+k}+\xi_p),
\label{noneq}
\end{eqnarray}
which can be easily derived from (\ref{epsilon}) and
(\ref{eta2}) respectively.
Substituting these expressions into the equation (\ref{kinetic1}) we
arrive at the electron-electron collision integral for the degenerate plasma:
\begin{eqnarray}
I_{ee}=\int\frac{d^3k}{(2\pi)^3}\frac{d^3p'}{(2\pi)^3}
\bigg(\frac{4\pi e^2}{k^2}\bigg)^2\frac{8\pi}
{|\epsilon(\xi_{p+k}-\xi_p,k)|^2}
\delta(\xi_{p'+k}+\xi_{p-k}-\xi_{p'}-\xi_p)\times
\nonumber \\
\times\big[n_{p'+k}n_{p-k}(1-n_{p'})(1-n_p)-n_{p'}n_p(1-n_{p'+k})(1-n_{p-k})\big]
\label{ee}
\end{eqnarray}
Thus the kinetic equation can be written in a standard form
\begin{equation}
\frac{dn_p}{dt}=I_{eph}+I_{ee},
\end{equation}
where the collision integrals $I_{eph}$ and $I_{ee}$ are defined
respectively by eqs. (\ref{eph1}) and (\ref{ee}).

In order to estimate the characteristic electron scattering time
we have to substitute the function
$n_p+\delta n_p$ instead of $n_p$ in the collision integral (\ref{kinetic1}).
The inverse inelastic scattering time $1/\tau_i$ is then defined as
a coefficient in front of the term  $\delta n_p$ describing deviations
from equilibrium. Making use of an obvious identity
$1-2n_{p-k}=\tanh ((\xi_p-\omega)/2T)$ we get
\begin{eqnarray}
\frac{1}{\tau_i(p)}&=&
%\frac{e^2}{\pi^2}\int d\omega d^3k
%{\rm Im}\left(\frac{-1}{k^2\epsilon(\omega,k)}\right)
%\delta(\omega+\xi_{p-k}-\xi_p)\bigg(
%\coth\frac{\omega}{2T} + 1-2n_{p-k}\bigg)
%\nonumber \\
%&=&
\frac{e^2}{\pi^2}\int d\omega d^3k
{\rm Im}\left(\frac{-1}{k^2\epsilon(\omega,k)}\right)
\delta(\omega+\xi_{p-k}-\xi_p)\bigg(
\coth\frac{\omega}{2T} +\tanh\frac{\xi_p-\omega}{2T} \bigg).
\label{ti}
\end{eqnarray}
It is clear from this equation that the time $\tau_i$ becomes infinite
at zero temperature and at the Fermi energy due to the Pauli principle.
The same is true for
the inelastic scattering time due to electron-phonon interaction. Thus
within the same general formalism we have demonstrated a clear
distinction between the decoherence time $\tau_{\varphi}$ and the
inelastic scattering time $\tau_i$. The latter becomes infinite
at $T=0$ because the excitation due to noise ($\coth(\omega/2T)$) and
the energy losses due to radiation ($\tanh((\xi-\omega)/2T)$, see the
discussion below) exactly compensate each other. This is in the
agreement with the Pauli principle: electrons cannot change their
energy because at $T=0$ all states below the Fermi level are occupied.
At the same time the value $\tau_{\varphi}$ remains finite even
at $T=0$ because this time is sensitive to random noise of the environment
only (be it classical or quantum). The effect of such noise leads to
decoherence.  We will return to this discussion below.

The above kinetic equations were derived for a simple case of
a clean system and do not account for the effect of elastic
scattering. In the case of a disordered
metal the electron momentum is not conserved and the
whole derivation becomes much more complicated. One can
demonstrate (see e.g. \cite{AAK1}) that in the diffusive
limit the result is roughly equivalent to a substitution
$$
\delta(\omega+\xi_{p-k}-\xi_p) \to
\mbox{Re}\bigg[\frac{1}{i\omega +Dk^2}\bigg]
$$
in the expression (\ref{ti}). An extended analysis of the
inelastic scattering time in various limits is given in Ref.
\onlinecite{Blanter}.

One can also formulate an alternative approach and derive
the quasiclassical Langevin equations describing electron
dynamics in a disordered metal. In doing so, we
follow the same procedure as one described in Refs.
\onlinecite{Schmid1,GZ92}.

Consider only close electron paths for which the values
$\bbox{r}^-=\bbox{r}_1-\bbox{r}_2$ and $\bbox{p}^-=\bbox{p}_1-\bbox{p}_2$
are small. Then we can expand the effective action in the
exponent of eq. (\ref{J}) in powers of $\bbox{r}^-$ and
$\bbox{p}^-$ keeping only the quadratic terms. The action
becomes Gaussian in terms of these variables and the integral
(\ref{J}) is dominated by the saddle point trajectories:
$\delta S/\delta \bbox{p}^-=0$ and $\delta S/\delta \bbox{r}^-=0$.
The first equation coincides with one without dissipation:
$\bbox{\dot r}=\bbox{p}/m$. With the aid of this equation the
momentum can be easily excluded and we get:
\begin{equation}
m \bbox{\ddot r} + \nabla U(\bbox{r}) + e^2\big[1-2n(\bbox{r},m\bbox{\dot
r})\big] \int\limits_{-\infty}^t dt'\nabla_r R(t-t',\bbox{r}(t)-\bbox{r}(t'))=
-e\bbox{E}(t,\bbox{r}).
\label{Langevin}
\end{equation}
Here $\bbox{E}(t,\bbox{r})$ is a fluctuating electric field. Equilibrium
fluctuations of this field are described by the correlator
\begin{equation}
\langle E_i(t_1,\bbox{r}_1)E_j(t_2,\bbox{r}_2)\rangle=
4\pi\delta_{ij}\int\frac{d\omega d^3k}{(2\pi)^4}
{\rm Im}\bigg(\frac{-\coth\frac{\omega}{2T}}{\epsilon(\omega,k)}\bigg)
e^{-i\omega(t_1-t_2)+ik(\bbox{r}_1-\bbox{r}_2)}
\label{corr}
\end{equation}
If needed, the generalization of (\ref{Langevin},\ref{corr}) to a strongly
nonequilibrium situation can be also provided. Also more general
expressions for the kernel $R(t,\bbox{r_1},\bbox{r_2})$ and for the correlation
function (\ref{corr}) for the case $|\bbox{r_1}-\bbox{r_2}|<l$ can be easily
derived. Combining these expressions with (\ref{Langevin}) one can obtain
a quasiclassical description of electron dynamics also at scales $\lesssim l$.
However such details are not important for us here, eq. (\ref{Langevin})
is presented merely to illustrate important physical effects.

The equation (\ref{Langevin}) obviously satisfies the requirement of
causality and captures all the essential features of electron dynamics
in a metal. E.g. it demonstrates that electrons in a metal cannot
infinitely decrease their energy: effective damping due to the
presence of the environment (described by the last term in
the left hand side of (\ref{Langevin})) is zero at the Fermi energy
($n=1/2$) and becomes negative below this energy.
Thus electrons with the initial energy above $\mu$ will loose
it before they reach the Fermi level. On the contrary, holes
with the initial energy below $\mu$ will be pushed up to the Fermi surface.
This simple example demonstrates again that our analysis accounts
explicitely for the Pauli principle. The corresponding information
is contained in the influence functional which depends on the occupation
numbers.

The damping term in eq. (\ref{Langevin}) depends on the
function $R(t,\bbox{r})$ which is determined by the correlation
function $\langle V^+V^-\rangle$ (see (\ref{VV1})). The physical
origin of this damping term is quite transparent: the electron (or
the hole) propagating in a metal produces the screened electric
potential due to the presence of other electrons and ``feels''
this potential itself.  In this sence eq. (\ref{Langevin}) is
similar to the equation of motion of a high energy particle (e.g. muon)
in a metal. The important difference between these two cases, however,
lies in the factor $1-2n$ which is present in our case due to the Pauli
principle. Formally this factor enters due to fluctuations of the field
$V^-$ which is ``sensitive'' to the Pauli principle. The fluctuating electric
field $E$ in the right hand side of eq. (\ref{Langevin}) is, on the
contrary, not affected by the Pauli principle because its correlation
function depends only on the field $V^+$.

With the aid of eq. (\ref{Langevin}) it is also easy to understand why
the real part of the influence functional $S_R$
(\ref{SR}) does not contribute to the decoherence time. According to
(\ref{Langevin}) the phase
difference acquired by the electron propagating along some
classical path can be split into two parts: the
regular contribution due to damping ($S_R$) which depends only on the electron
trajectory, and irregular part due to noise ($S_I$). Considering
now the contribution from a pair of time reversed paths, we observe that
the regular contributions are the same and cancel each other because
they enter with a different sign. Only irregular contributions due to
noise survive and determine $\tau_\varphi$. For each path
the regular contribution may have a different value depending on the
path and energy and even vanish (for energies at the Fermi level).
However by no means this affects the noise terms and thus $\tau_{\varphi}$
which always remains finite.

\section{Discussion}

With the aid of a general formalism of Green-Keldysh functions we
have analyzed a fundamental effect of
quantum decoherence of the electron wave function in a disordered
metal due to interaction. Our treatment was carried out with no
more assumptions than the usual ones in the weak localization theory:
the elastic mean free path
was considered large as compared to the Fermi wavelength $p_Fl \gg 1$
and interactions were assumed to be sufficiently weak. Actually the
main effect does not really depend even on these general assumptions,
it can be observed already from a formally exact equation for the
density matrix (\ref{rho10},\ref{rho5}).

We have demonstrated that the effect of interaction of the electron
with other electrons and lattice ions in a disordered metal
{\it is equivalent} to the effect of a {\it dissipative} environment.
This effect is similar (although not exactly equivalent) to that of
a dissipative Caldeira-Leggett bath. Fluctuations in the dissipative
environment play the key role in the effect of quantum decoherence \cite{VA}.

Although the environment is dissipative it is {\it not} inelastic
scattering processes with energy transfer between real quantum
states that cause quantum decoherence at low $T$.
Rather this is the effect of {\it quantum noise in a dissipative environment}.
Due to interaction with quantum fluctuations of this environment the
electron not only ``goes'' through virtual states with different
energies but after a finite time $\tau_{\varphi}$ also looses
information about its initial phase.
At the same time no real processes with energy transfer exist at $T=0$.
It is also interesting to note that -- as can be seen e.g. from
eq. (\ref{VV}) -- voltage fluctuations and therefore the decoherence
rate $1/\tau_{\varphi}$ vanish in both limits $\sigma \to 0$ and
$\sigma \to \infty$.

The presence of Feynman paths which strongly deviate from the
classical paths depicted in Fig. 1a is also crucially important
for the whole effect. In the
case of disordered metals these are time reversed paths (Fig. 1b).
In the case of Caldeira-Leggett models other paths (e.g. instantons,
see \cite{CBM,Schmid2,SZ}) play a similar role. If the important
paths go sufficiently far from each other, electrons gain an
additional phase due to the environmental noise. This phase always
vanishes for the classical paths of Fig. 1a.

We clarified the role of the Pauli principle in the
effect of quantum decoherence. Although the Pauli principle
plays a crucial role in the inelastic electron-electron
collisions it does not affect quantum decoherence. The physical
reason for that is transparent: fluctuations of the collective
electric field in a metal are equivalent to local
fluctuations of the Fermi energy or the electron density
(see Section 2). Such fluctuations are essentially the same
for many electron states, and the Pauli principle does not play
any role.

On a very simple level the decoherence effect at low $T$
can be understood as follows. During interaction with other
electrons the electron phase changes. These
``other'' electrons (or some of them) ``go out of the game''
(the system is open!) and after some time the information
about the initial electron phase is lost. The same mechanism
works for a particle interacting with the Caldeira-Leggett bath.

We would like to emphasize that no special assumptions about
the ``open system'' were made in our analysis. We just used
the standard formalism of Green functions applicable
to {\it any} grand canonical ensemble in condensed matter
physics. Already by fixing the chemical potential $\mu$ and
allowing the total number of particles to vary one defines
the system as open. Thus our results are general and
in principle can be applied to any disordered system perhaps
except for ultrasmall objects with fixed numbers of particles.
The latter appear to be irrelevant for weak localization
measurements.

Our analysis provides no support for the point of view according
to which the effect of high frequency (quantum) fluctuations
of the electronic environment on $\tau_{\varphi}$ should be
detemined by {\it inelastic} processes with high energy transfers
and therefore should be obtained from the kinetic analysis.
The kinetic treatment deals with paths of Fig. 1a and
allows to calculate only $\tau_i$ but not
$\tau_{\varphi}$. In contrast, the latter is determined by
time reversed paths of Fig. 1c for all relevant frequencies.
In this sence there is no qualitative difference between the
effect of classical and quantum fluctuations of the
environment  on $\tau_{\varphi}$.
Both should be treated on equal footing.

There is also
no evidence that the Pauli principle ``forbids'' decoherence
at low $T$. Therefore the suggestion to account for the Pauli
principle by introducing an additional term proportional to
$\tanh (\omega /2T)$ into the expression for $\tau_{\varphi}$
does not appear to be justified. We have demonstrated that
the combination ``$\coth - \tanh$'' enters the expression
for $1/\tau_i$ (\ref{ti}) but {\it not} for the inverse decoherence
time $1/\tau_{\varphi}$. The latter contains only ``$\coth$''
and does not vanish at $T \to 0$.

We believe that the same should hold for a sometimes
conjectured cancellation of Keldysh diagrams at $T=0$.
To the best of our knowledge a rigorous formulation of the
diagramatic  representation for $\tau_{\varphi}$
is still lacking in the literature. An attempt to provide this
formulation was made by Fukuyama and Abrahams \cite{FA}. But it is
well understood by now \cite{Blanter}
that only diagrams relevant for the inverse inelastic time $1/\tau_i$
were taken into account in \cite{FA}. Cancellation of these
diagrams at $T=0$ is quite natural but it does not yet tell
anything about $\tau_{\varphi}$. Since our analysis should in
principle include {\it all} diagrams we see no reason to expect any
cancellation of diagrams for $\tau_{\varphi}$ even at $T=0$.

The low temperature saturation of $\tau_{\varphi}$ can cause
{\it dramatic} consequences for the existing picture of strong
localization \cite{T,A} in lower dimensions. There seems to be
no reason to challenge this picture as long as one neglects
Coulomb interaction between electrons. But -- as was demonstrated
above -- it is exactly this interaction that leads to quantum
decoherence at $T=0$. If one compares the effective
decoherence length $L_{\varphi}$ with the localization length
$L_{loc}$ one immediately sees that for the parameters of a typical metal
the former is {\it always} smaller than the latter. E.g. in 1d
our analysis yields $L_{\varphi} \sim \sqrt{D\tau_{\varphi}}
\sim l\sqrt{N_{ch}}$. This expression is {\it parametrically}
smaller than the localization length $L_{loc} \sim lN_{ch}$
for large number of conducting channels. Analogously in 2d
at low $T$ the weak localization correction to conductivity saturates
at the level $\delta \sigma_2 \sim -(e^2/2\pi^2)\ln (p_F^2al)$ and
the crossover from weak to strong localization never occurs.
In other words, all that implies that (at least for $p_Fl \gg 1$)
due to electron-electron
interaction the electron will loose coherence already before it can
get localized, i.e. {\it strong localization does not take place at all}
and the 1d and 2d metals {\it do not become insulators even at} $T=0$.

We can also add that it would be interesting to investigate the effect of
electron-electron interaction on more sophisticated weak localization
corrections, like ones due to quantum interference of diffusons
\cite{Hikami}. Here the point is that (unlike Cooperons) diffusons
do not decay at $T=0$ even in the presence of interaction. Nevertheless,
we cannot exclude that even at $T=0$ quantum
interference of diffusons can be also destroyed due to the same mechanism of
quantum noise in a dissipative electronic environment as it was
discussed above for Cooperons. This problem will be studied elsewhere.

Irrespectively to such details it is quite clear that an adequate description
of electron-electron interaction can seriously change the existing
picture of strong localization in low dimensional disordered metals.
This perspective is very interesting but not very surprizing. In fact,
the important role of electron-electron interaction was already pointed
out long ago by Finkelstein \cite{Fin}.

One more comment concerns the relation between the low temperature saturation
of the decoherence time $\tau_{\varphi}$ and the applicability of the
Fermi liquid theory. It is sometimes believed that the applicabitity
condition for the latter is $T\tau_{\varphi} \gg 1$. From a practical
point of view this condition is not very restrictive: e.g. for typical
experimental values of Ref. \cite{Webb}
the value $\tau_{\varphi}$ was found to be in the range $10^{-1} \div 10$ ns
and the above condition can be violated only at temperatures
below $1\div 100$ mK. However theoretically it is very important to
understand if a finite decoherence time at $T \to 0$ would imply
the breakdown of the Fermi liquid theory at low $T$. We believe
that there is no direct relation between $\tau_{\varphi}$ and the
lifetime of quasiparticles. The latter should be rather defined by
some inelastic time (e.g. by $\tau_i$) which is always much longer
than $\tau_{\varphi}$ at low $T$. Thus it appears that even for
$T\tau_{\varphi} <1$ there is still no evidence for the breakdown of
the Fermi liquid theory in metals. Previously an analogous conclusion
was reached in Ref. \onlinecite{AW}.

Finally, we briefly discuss the agreement between our results and the
available experimental data.
The comparison between theoretical and experimental
values for the decoherence length $L_\varphi=\sqrt{D\tau_\varphi}$
at zero temperature is given in the table \ref{table}.
To calculate
$L_\varphi$ we first estimate the decorence time with the aid of (\ref{tau2}).
At $T=0$ the time $\tau_\varphi$ can be conveniently
expressed in terms of  measurable quantities:
\begin{equation}
\tau_\varphi=\sqrt{6}\frac{R_qL}{Rv_F},
\label{tauexp}
\end{equation}
where $R_q=\pi/2e^2=6453$  $\Omega$ is a quantum resistance, $L$ is the wire
length and $R$ is the total resistance. The Fermi velocity for
gold wires was taken to be \cite{Kittel} $1.39\times 10^6$ m/s.
The diffusion coefficient $D$ was estimated with the aid of the
Drude formula $\sigma=2e^2N_0D$. The density of states for gold is chosen
to be \cite{Kittel} $N_0=6\times 10^{12}$ s/m$^3$. Note that
the numerical values for $D$ are not identical to those given in \cite{Webb}.

\begin{table}[h]
\caption{Theoretical values of the coherence length
$L_\varphi=\sqrt{D\tau_\varphi}$ in comparison with the
experimental results \cite{Webb}}
\begin{tabular}{cddddddd}
 Sample & $w$, nm & $t$, nm & $L$, $\mu$m &
 $R/L$, $\Omega/\mu$m &
D, $10^{-3}$ m$^2$/s & $L_\varphi^{\rm exp}$, $\mu$m &
$L_\varphi^{\rm theor}$, $\mu$m\\
 \hline
 Au-1 & 60  & 25 & 57.9 & 29.14 & 7.8  & 5.54 & 1.8    \\
 Au-2 & 110 & 60 & 207  & 1.46  & 35.5 & 16   & 16.5   \\
 Au-3 & 100 & 35 & 155  & 9.31  & 10.5 & 5.2  & 3.6    \\
 Au-4 & 60  & 25 & 57.9 & 31.29 & 7.3  & 3.6  & 1.6     \\
 Au-5 & 190 & 40 & 18.9 & 191.7 & 0.24 & 0.35 & 0.12    \\
 Au-6 & 180 & 40 & 155  & 2.91  & 16.3 & 8    & 8.1     \\
\end{tabular}
\label{table}
\end{table}

The width, the thickness and the length of the wire are denoted
respectively by $w$, $t$ and $L$; $R/L$ is the resistance per unit length and
$L_\varphi^{\rm exp}$, $L_\varphi^{\rm theor}$ are experimental and
theoretical values of the decoherence length. The agreement between
both looks very reasonable for all samples, especially if one
takes into account an uncertainty in a numerical prefactor in our
formulas due to the cutoff procedure and possible effects of the
sample geometry.  It is also important to emphasize that our comparison
involves {\it no} fitting parameters.

\begin{figure}[h]
\centerline{\psfig{file=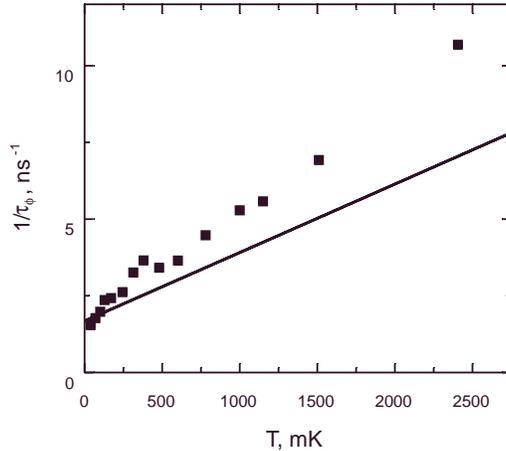,height=6cm}}
\caption{The dependence of the decoherence rate $1/\tau_\varphi$
on temperature for one of the thin gold wires \cite{Webb2}.
The wire has the following parameters: $w=35$ nm, $t=20$ nm, $L=231$ $\mu$m,
$R=4411$ $\Omega$, $D=2.56\times 10^{-2}$ m$^2$/s.
Experimental points are shown by squares, the solid curve represents the
teoretical dependence (\ref{tau2}).
The experimental data were kindly provided by P. Mohanty. }
\label{plot}
\end{figure}

The temperature dependence of the inverse decoherence time $1/\tau_\varphi$ is
plotted in Fig. \ref{plot}. The agreement between theoretical and
experimental results is again reasonable. It appears, that within
the temperature interval of Fig. \ref{plot} temperature dependence
of the experimental data is not far from a linear one.
Note, that the temperature
dependent part of eq. (\ref{tau2}) is sensitive to details of the
low frequency cutoff. The plotted theoretical dependence corresponds to
the lower cutoff frequency equal to $1/\tau_\varphi$ at $T=0$. With
a slight adjustment of this cutoff one can in principle reach a
perfect agreement with the experimental data. However, we believe
that this adjustment is not needed: within the accuracy of our
calculation the agreement is already very good.

Although we do not present the comparison of our theoretical
results with the experimental data from Refs. \onlinecite{Gio,Pooke,Hir}
we note that all these data are sufficiently well described by our formulas
(\ref{tau2},\ref{11}). The agreement with the experimental results
by Mueller {\it et al.} \cite{Mueller} for 2d gold films turns out to be
worse: the value $\tau_\varphi$ estimated from eq. (\ref{11}) is
considerably shorter than the experimentally measured value. The reason for
this discrepancy is not clear at the moment. We can speculate that it
might be due to the high purity of the films \cite{Mueller} which
can make the diffusive approximation insufficient. If this is the case
the resistance of the films \cite{Mueller} is rather due to
electron scattering at the boundaries
and our theoretical analysis should be substantially modified. This
problem will not be discussed here.

In all the above (and many other) experiments no crossover from weak
to strong localization was not detected even at the lowest temperatures
$20\div40$ mK. Very recently in Ref. \cite{Gersh}
a sharp increase of the resistance by several orders of magnitude
was observed in quasi-1d GaAs structures with small number of
conducting channels $N_{ch} \sim 10$. This increase of the resistance
was interpreted as a crossover from weak to strong localization.
Although this could be one possible scenario we believe
the interpretation of the above effect is still an open problem.
Indeed, if one estimates the crossover temperature from the standard
criterion $L_{\varphi} \sim L_{loc}$ and makes use of (\ref{AAK})
for the parameters \cite{Gersh} one arrives at the crossover temperature
$T \sim 1\div 10$ mK. However the increase of the resistance was
observed at a much higher temperature $T\gtrsim 1$ K. This discrepancy
might indicate the important role of Coulomb effects in the system.
Analogously Coulomb effects
were found to play the key role in many experimens with
disordered 2d systems, see e.g. \cite{Wash} for the corresponding
discussion. Further experiments are definitely needed for better
understanding of the electron-electron interaction effects in
disordered low dimensional systems.

We would like to thank Ya. Blanter, G. Blatter, C. Bruder, D. Geshkenbein,
V. Kravtsov, A. Mirlin, A. van Otterlo, M. Paalanen, A. Schmid, G. Sch\"on
and P. W\"olfle for valuable discussions, comments and/or encouragement.
We are also indebted to the authors \cite{Webb2} for communication
of their new results prior to publication.
This work was supported by the Deutsche Forschungsgemeinschaft within
SFB 195 and by the INTAS-RFBR Grant No. 95-1305.

\newpage

\appendix
\section{Density matrix}

With the aid of eq. (\ref{G1}) the time integral in the last term of
the Dyson equation (\ref{Dyson}) can be transformed as follows:
\begin{eqnarray}
\int\limits_0^t dt' \hat G_0(t_1,t')e\hat V(t')\hat G(t',t_2)
%&=&
%\int\limits_0^t dt' \hat G_0(t_1,t')
%\bigg[\hat 1\delta(t'-t_2)-
%\left(i\frac{\partial}{\partial t'}-\hat H_0(t')\right)\hat G(t',t_2)\bigg]
%\nonumber \\
&=&\hat G_0(t_1,t_2) -
\int\limits_0^t dt' \hat G_0(t_1,t')
\left(i\frac{\partial}{\partial t'}-\hat H_0(t')\right)\hat G(t',t_2)
\nonumber \\
&=& \hat G_0(t_1,t_2)-i\hat G_0(t_1,t)\hat G(t,t_2)
+i\hat G_0(t_1,0)\hat G(0,t_2) +
\nonumber \\
&&
\int\limits_0^t dt'
\left(i\frac{\partial}{\partial t'}+\hat H_0(t')\right)\hat G_0(t_1,t')
\hat G(t',t_2)
\nonumber \\
&=& \hat G_0(t_1,t_2)-\hat G(t_1,t_2)-i\hat G_0(t_1,t)\hat G(t,t_2)
+i\hat G_0(t_1,0)\hat G(0,t_2)
\label{int1}
\end{eqnarray}
Here we performed the integration by parts over the time $t'$
and made use of the equation
$$
\left(i\frac{\partial}{\partial t'}+\hat H_0(x')\right)
\hat G_0(t_1,t')=-\delta(t_1-t')\delta(x_1-x'),
$$
which defines the field-free Green-Keldysh function $\hat G_0$.
Substituting the result (\ref{int1}) into eq. (\ref{Dyson}) we arrive
at (\ref{diG}).

Let us substitute the representation (\ref{Uf}) into (\ref{diG}). Then we find
\begin{equation}
\hat U_0(t_1,t)(-\hat b+\hat f_0(t))\hat U_V(t,t_2) (\hat a+\hat f_V(t_2))-
\hat U_0(t_1,0)(\hat a+\hat f_0(0))\hat U_V(0,t_2) (-\hat b+\hat f_V(t_2)) = 0
\label{byparts}
\end{equation}
It is now convenient to represent the operator $\hat f_V$ in the form
\begin{equation}
\hat f_V(t_2)=\hat U_V(t_2,0)\hat g(t_2)\hat U_V(0,t_2).
\label{g}
\end{equation}
Then for $\hat g$ we get
\begin{equation}
[1+(\hat b-\hat f_0(0))(\hat U_0(0,t)\hat U_V(t,0)-1)]\hat g(t_2)=
\hat f_0(0) - (\hat b-\hat f_0(0))(\hat U_0(0,t)\hat U_V(t,0)-1)\hat a.
\label{g1}
\end{equation}
Let us put $t_2=t$ and introduce the scattering
matrix $S=U_0(0,t)U_V(t,0)=s_1\hat a+s_2\hat b$. This matrix is diagonal
because both $U_0$ and $U_V$ are the diagonal matrices. The matrix elements
$s_1$ and $s_2$ are defined as $s_{1,2}=u_0(0,t)u_{1,2}(t,0)$. Making use of
the above notations and rewriting the equation (\ref{g1}) in components we find
\begin{eqnarray}
\left[\left(\begin{array}{cc}
             1 & 0 \\
             0 & 1
           \end{array}\right)
+
\left(\begin{array}{cc}
       \rho_0 & -\rho_0 \\
       -1+\rho_0 & 1-\rho_0
     \end{array}\right)
\left(\begin{array}{cc}
      s_1-1 & 0 \\
        0  & s_2-1
      \end{array}\right)\right]
\left(\begin{array}{cc}
      g_{11} & g_{12}\\
      g_{21} & g_{22}
     \end{array}\right) =
\nonumber \\
=
\left(\begin{array}{cc}
       -\rho_0 & \rho_0 \\
       1-\rho_0 & \rho_0
     \end{array}\right)
-
\left(\begin{array}{cc}
       \rho_0 & -\rho_0 \\
       -1+\rho_0 & 1-\rho_0
     \end{array}\right)
\left(\begin{array}{cc}
      s_1-1 & 0 \\
        0  & s_2-1
      \end{array} \right)
\left(\begin{array}{cc}
      1 & 0 \\
      0 & 0
   \end{array}\right)
\label{matrix}
\end{eqnarray}
Multiplying matrices
%\begin{equation}
%\left(\begin{array}{cc}
%    1+\rho_0(s_1-1) & -\rho_0(s_2-1) \\
%   -s_1+1+\rho_0(s_1-1) & s_2-\rho_0(s_2-1)
%   \end{array}\right)
%\left(\begin{array}{cc}
%     g_{11} & g_{12} \\
%     g_{21} & g_{22}
%     \end{array}\right)=
%\left(\begin{array}{cc}
%     -\rho_0s_1 & \rho_0 \\
%    s_1-\rho_0s_1 & \rho_0
%    \end{array}\right)
%\label{matrix1}
%\end{equation}
and keeping only the part of the resulting matrix equation which depends on
$g_{12}$ and $g_{22}$ we obtain
\begin{equation}
\left(\begin{array}{cc}
    1+\rho_0(s_1-1) & -\rho_0(s_2-1) \\
   -s_1+1+\rho_0(s_1-1) & s_2-\rho_0(s_2-1)
   \end{array}\right)
\left(\begin{array}{c}
      g_{12} \\
      g_{22}
     \end{array}\right)=
\left(\begin{array}{c}
      \rho_0 \\
     \rho_0
    \end{array}\right)
\label{matrix2}
\end{equation}
Substracting the second equation from the first one we get
$s_1g_{12}-s_2g_{22}=0$, or $g_{22}=s_2^{-1}s_1g_{12}$. Subsituting
this result into the first equation we find
%\begin{equation}
%\big[1+\rho_0(s_1-1)-\rho_0(s_2-1)s_2^{-1}s_1\big]g_{12}=\rho_0.
%\end{equation}
%After obvious simplifications we get
\begin{equation}
\big[1+\rho_0(s_2^{-1}s_1-1)\big]g_{12}=\rho_0.
\end{equation}
Note that the S-matrices enter the above equation only in the combination
$s_2^{-1}s_1$ which does not contain the evolution operator $u_0$, i.e.
we have $s_2^{-1}s_1=u_2(0,t)u_1(t,0)$ and
\begin{equation}
[1-\rho_0(0)(1-u_2(0,t)u_1(t,0))]g_{12}(t)=\rho_0(0).
\label{g12}
\end{equation}
Rewriting (\ref{g}) for the 1,2-component of the matrix $\hat f_V(t)$
$$
f_{12}(t)=u_1(t,0)g_{12}(t)u_2(0,t),
$$
and making use of the identity $\rho_V(t)=f_{12}(t)$ we arrive at the
result (\ref{rho}).

\section{Effective action and FDT}

For $V_{1,2}$ the electron Green functions $G_{12}$ and $G_{21}$ can be
expressed in the form
\begin{equation}
G_{12}=iu_0(t_1,0)\rho_0u_0(0,t_2),
\qquad G_{21}=-iu_0(t_1,0)(1-\rho_0)u_0(0,t_2).
\label{G12}
\end{equation}
In thermodynamic equilibrium we have $u_0(t,0)\rho_0u_0(0,t)=\rho_0$ for
any time $t$.

Let us introduce the basis of the eigenfunctions for the single
electron Hamiltonian, $H_0\psi_k=\xi_k\psi_k$. Without loss of generality
we can choose these eigenfunctions to be real. The initial
density matrix $\rho_0$ is assumed to be diagonal in the basis $\psi_k$, namely
$\rho_0=\sum_k n_k|\psi_k\rangle\langle\psi_k|$.
This assumption is justified only for weakly interacting particles.
Then the functions $G_{12}$, $G_{21}$ (\ref{G12}), $G^R$ and $G^A$
(\ref{GRA}) can be written in the form
\begin{eqnarray}
G_{12}(t_1,t_2,\bbox{r}_1,\bbox{r}_2)=i\sum\limits_k e^{-i\xi_k(t_1-t_2)}n_k
\psi_k(\bbox{r}_1)\psi_k(\bbox{r}_2),
\nonumber \\
G_{21}(t_1,t_2,\bbox{r}_1,\bbox{r}_2)=-i\sum\limits_k e^{-i\xi_k(t_1-t_2)}(1-n_k)
\psi_k(\bbox{r}_1)\psi_k(\bbox{r}_2),
\nonumber \\
G^R(t_1,t_2,\bbox{r}_1,\bbox{r}_2)=-i\theta(t_1-t_2)
\sum\limits_k e^{-i\xi_k(t_1-t_2)}
\psi_k(\bbox{r}_1)\psi_k(\bbox{r}_2),
\nonumber \\
G^A(t_1,t_2,\bbox{r}_1,\bbox{r}_2)=i\theta(t_2-t_1)
\sum\limits_k e^{-i\xi_k(t_1-t_2)}\psi_k(\bbox{r}_1)\psi_k(\bbox{r}_2).
\end{eqnarray}
With the aid of these expressions the kernels (\ref{chi1},\ref{eta1})
can be represented as follows
\begin{equation}
\chi(t,\bbox{r}_1,\bbox{r}_2)= 2ie^2\theta(t)\sum\limits_{k,q}
e^{-i(\xi_k-\xi_q)t}(n_k-n_q)\psi_k(\bbox{r}_1)\psi_k(\bbox{r}_2)
\psi_q(\bbox{r}_2)\psi_q(\bbox{r}_1);
\label{chi}
\end{equation}
\begin{equation}
\eta(t,\bbox{r}_1,\bbox{r}_2)=
\frac{1}{2}e^2\sum\limits_{k,q}
e^{-i(\xi_k-\xi_q)t}
[n_k(1-n_q)+n_q(1-n_k)]
\psi_k(\bbox{r}_1)\psi_k(\bbox{r}_2)\psi_q(\bbox{r}_2)\psi_q(\bbox{r}_1).
\label{eta}
\end{equation}
Performing the Fourier transformation in time:
\begin{equation}
\chi(\omega,\bbox{r}_1,\bbox{r}_2)=
-2e^2\sum\limits_{k,q}
\frac{n_k-n_q}{\omega-\xi_k+\xi_q+i0}
\psi_k(\bbox{r}_1)\psi_k(\bbox{r}_2)\psi_q(\bbox{r}_2)\psi_q(\bbox{r}_1),
\label{chi2}
\end{equation}
\begin{equation}
\eta(\omega,\bbox{r}_1,\bbox{r}_2)=
\pi e^2\sum\limits_{k,q}
\delta(\omega-\xi_k+\xi_q)
[n_k(1-n_q)+n_q(1-n_k)]
\psi_k(\bbox{r}_1)\psi_k(\bbox{r}_2)\psi_q(\bbox{r}_2)\psi_q(\bbox{r}_1),
\label{eta2}
\end{equation}
and substituting the equilibrium distribution function
$n_k=1/(e^{\xi_k/T}+1)$ one immediately arrives at (\ref{FDT}).

%\end{multicols}


\begin{references}
\bibitem{FN0} Strictly speaking this procedure is correct only for
times $t \sim \tau_{\varphi}$. At $t \gg \tau_{\varphi}$ the low frequency
cutoff in the integral (\ref{tau1}) should be fulfilled at $1/t$. In 2d and 3d
systems this does not lead to any changes in the results. The same is
true for 1d systems at low temperatures  $T<T^{(1)}_q$. At higher $T$ the
decay of correlations in time will be even faster than exponential ($ \propto
\exp (-(t/\tau_{\varphi})^{3/2})$ -- see e.g. \cite{Imry}), but again the
characteristic time scale is defined by $\tau_{\varphi}$.
\end{references}
\end{document}